\documentclass[a4paper,12pt]{article}

\usepackage{amsmath,amssymb,amsfonts}

\newlength{\xtrawidth}
\newlength{\xtraheight}
\newcommand{\beq}{\begin{equation}}
\newcommand{\eeq}{\end{equation}}
\newcommand{\ba}{\begin{array}}
\newcommand{\ea}{\end{array}}
\newcommand{\bt}{\begin{tabular}}
\newcommand{\et}{\end{tabular}}
\newcommand{\bea}{\begin{eqnarray}}
\newcommand{\eea}{\end{eqnarray}}
\newcommand{\bean}{\begin{eqnarray*}}
\newcommand{\eean}{\end{eqnarray*}}
\newcommand{\eref}[1]{(\ref{#1})}
\newcommand{\tref}[1]{Table~\ref{#1}}

\newcommand{\nn}{\nonumber}
\newcommand{\comment}[1]{}

\newtheorem{theorem}{\sf THEOREM}

\newcommand{\rk}{\mathop{{\rm rk}}}
\newcommand{\ind}{\mathop{{\rm ind}}}

\newcommand{\coker}{\mathop{{\rm coker}}}
\newcommand{\im}{\mathop{{\rm im}}}
\newcommand{\IC}{\mathbb{C}}
\newcommand{\IP}{\mathbb{P}}
\newcommand{\IZ}{\mathbb{Z}}
\newcommand{\cO}{{\cal O}}

\newcommand{\cN}{{\cal N}}
\newcommand{\cV}{{\cal V}}
\newcommand{\cB}{{\cal B}}
\newcommand{\cC}{{\cal C}}
\newcommand{\cK}{{\cal K}}
\newcommand{\cW}{{\cal W}}

\def\fnote#1#2{\begingroup\def\thefootnote{#1}\footnote{#2}
     \addtocounter{footnote}{-1}\endgroup}

\begin{document}

\title{{\LARGE Heterotic Compactification, An Algorithmic Approach\\
%    Stable Monads and Computer Algebra}
}}
\author{
Lara B. Anderson${}^{1,3}$,
Yang-Hui He${}^{1,2}$,
Andr\'e Lukas${}^{3}$
}
\date{}
\maketitle
\begin{center}
{\small
${}^1$ {\it Mathematical Institute, Oxford University, 
\\ $~~~~~$ 24-29 St.~Giles', Oxford OX1 3LB, U.K.}\\
% ${}^2$ {\it Magdalen College, Oxford, OX1 4AU, U.K.}\\
${}^2${\it Merton College, Oxford, OX1 4JD, U.K.}\\
${}^3${\it Rudolf Peierls Centre for Theoretical Physics, Oxford
  University,\\
$~~~~~$ 1 Keble Road, Oxford, OX1 3NP, U.K.}\\
% ${}^5${\it Balliol College, Oxford, OX1 3BJ, U.K.}\\
\fnote{}{anderson@maths.ox.ac.uk}
\fnote{}{hey@maths.ox.ac.uk} 
\fnote{}{lukas@physics.ox.ac.uk}
}
\end{center}

\abstract{We approach string phenomenology from the perspective of
  computational algebraic geometry, by providing new and efficient
  techniques for proving stability and calculating particle spectra in
  heterotic compactifications. This is done in the context of complete
  intersection Calabi-Yau manifolds in a single projective space where
  we classify positive monad bundles. Using a combination of analytic
  methods and
  computer algebra we prove stability for all such bundles and compute the
  complete particle spectrum, including gauge singlets. In particular,
  we find that the number of anti-generations vanishes for all our
  bundles and that the spectrum is manifestly moduli-dependent.}

\newpage
\tableofcontents

%
%==========
%
\section{Introduction}

Compactification of the $E_8\times E_8$ heterotic string on Calabi-Yau
three-folds~\cite{Candelas:1985en,greene} is one of the oldest approaches to
particle phenomenology from string theory. Heterotic models have a
number of phenomenologically attractive features typically not shared
by alternative string constructions. Most notably,
gauge unification is ``automatic'' and standard model families
originate from an underlying spinor-representation of $\rm{SO}(10)$.
However, despite its long history and substantial recent
progress~\cite{Braun:2005ux}--\cite{Braun:2005zv},
heterotic model building is still a long way away
from one of its major goals: finding an example which does not
merely have standard model spectrum but reproduces the standard model exactly,
including detailed properties such as, for example, Yukawa couplings.

One of the main obstacles in achieving this goal is the inherent
mathematical difficulty of heterotic models. In addition to a
Calabi-Yau three-fold $X$, heterotic models require two holomorphic
(semi)-stable vector bundles $V$ and $\tilde{V}$ on $X$. Except for
the simple case of standard embedding, where $V$ is taken to be the
tangent bundle $TX$ of the Calabi-Yau space and $\tilde{V}$ is
trivial, construction of these vector bundles is often not
straightforward and the computation of their properties is usually
involved.  For example, stability of these bundles, an essential
property if the model is to preserve supersymmetry, is notoriously
difficult to prove. In addition, when searching for realistic particle
physics from heterotic string theory, these mathematical obstacles
have to be resolved for a large number of Calabi-Yau spaces and
associated bundles, as every single model (or even a small number of
models) is highly likely to fail when confronted with the detailed
structure of the standard model.  The main purpose of this paper is to
present an algorithmic approach to this problem by combining analytic
methods and computer algebra.  By an algorithmic approach we mean a
set of techniques which allow us to construct classes of vector
bundles on (certain) Calabi-Yau spaces systematically, prove their
stability and compute the resulting low-energy particle spectra
completely. In this paper we will focus on developing the necessary
computational methods by concentrating on the five Calabi-Yau
manifolds which can be obtained by intersections in an ordinary
projective space. A generalization of these methods to more general
complete intersection Calabi-Yau manifolds and a detailed analysis of
the particle physics properties of these models will be the subject of
forthcoming publications~\cite{us}.

Starting with the pioneering work in~\cite{Witten:1985bz,Distler:1987ee}, 
there has been continuing
activity on Calabi-Yau based non-standard embedding models
over the years. Recently, there has been significant
progress both from the mathematical and the model-building viewpoint,
leading to models edging closer and closer towards the standard
model~\cite{Braun:2005nv,Bouchard:2005ag}. 
Two types of constructions, one based on
elliptically fibered Calabi-Yau spaces with bundles of the
Friedman-Morgan-Witten type~\cite{spectral} and
generalizations~\cite{Braun:2005ux}--\cite{Braun:2005zv},
\cite{thomas}--\cite{Braun:2006ae}, the other based on complete
intersection Calabi-Yau spaces with monad
bundles~\cite{Distler:1987ee}, \cite{Blumenhagen:2006wj}--\cite{Douglas:2004yv},
have been pursued in the literature. In this paper, we will
work within the context of the second approach using complete
intersection Calabi-Yau manifolds and monad bundles. To explain our
motivation for this choice we remind the reader of the usual
``two-step'' symmetry breaking in heterotic models. In the first step,
the $E_8$ gauge group is broken to one of the standard grand unified
groups $E_6$, $\rm{SO}(10)$ or ${\rm SU}(5)$ by a bundle $V$ with
structure group $\rm{SU}(3)$, $\rm{SU}(4)$ or $\rm{SU}(5)$,
respectively.  Then a Wilson line is introduced to break further to
the standard model group (times possible $U(1)$ factors). This second
breaking requires a non-trivial first fundamental group of the
Calabi-Yau space $X$ which is normally achieved by dividing $X$ by a
discrete symmetry. For complete intersection Calabi-Yau manifolds,
this last procedure of dividing by a discrete symmetry group is greatly
facilitated by the presence of an ambient projective space.
This is one of our main motivations for working with this class of
manifolds, although analyzing discrete symmetries and Wilson line
breaking explicitly will be the subject of a future publication~\cite{us}.
Another major reason for our choice of models is that
all relevant objects can be readily described in the language
of commutative algebra and, therefore, lent themselves to an analysis
based on computer algebra.

In this paper, we will construct all positive monad bundles of rank 3,
4 and 5 on the five complete intersection Calabi-Yau spaces in a
single projective space subject to two additional constraints. First, the
bundles should be such that heterotic anomaly cancellation can be
accomplished and second, their chiral asymmetry should be a (non-zero)
multiple of three. We find $37$ examples in total. We then prove
stability for all these bundles using a variant of a simple criterion
due to Hoppe~\cite{hoppe}. Recently, this criterion has been
used~\cite{maria}, although in a slightly different way from the
present paper, to prove stability for a class of positive bundles on
the quintic~\cite{Douglas:2004yv}. Further, we compute the complete
spectrum for all
bundles, including gauge singlet fields. It turns out that a common
feature of our models is that they only lead to generations but no
anti-generations. While the present paper deals with a relatively
small number of examples, we have shown that the relevant methods can
be applied in a systematic and algorithmic way. We expect that a
significantly larger class of complete intersection Calabi-Yau spaces
and bundles on them can be treated in a similar way (see
\cite{Distler:2007av} for a recent constraint on classifying bundles
in general). This
generalization and the analysis of the particle physics of the
resulting models will be the subject of future work~\cite{us}.

The plan of the paper is as follows. In the next section, we will
briefly review the main general features of $E_8\times E_8$ heterotic
compactifications. In Section 3, we discuss the monad construction,
its main properties and prove a number of general results for such
bundles. In Section 4, we classify the positive monad bundles on our
five Calabi-Yau spaces, prove their stability and compute the
spectra. After our conclusions in Section 5, Appendix A follows with a
short summary
of the relevant tools in commutative algebra and how they are applied
in the context of the Macaulay computer algebra package~\cite{m2}. The
final Appendices contain several useful technical results.

%%
%--
%%
\section{Heterotic Compactification and Physical Constraints}
\label{s:het}
To set the scene, we would now like to briefly review the basic
structure of $E_8\times E_8$ heterotic vacua on Calabi-Yau three-folds 
(see Ref.~\cite{gsw,Donagi:2004ia,Braun:2005zv}).

In addition to a Calabi-Yau three-fold $X$ with tangent bundle, $TX$,
we need two holomorphic vector bundles $V$ and $\tilde{V}$ with
associated structure groups which are sub-groups of $E_8$. In the
present context, we will be interested in bundles with rank $n=3,4,5$
and corresponding structure group $G=\rm{SU}(n)$. In general,
heterotic vacua can also contain five-branes which appear as M
five-branes in the 11-dimensional strong-coupling limit and as NS
5-branes in the 10-dimensional weakly coupled theory. In either case,
for a supersymmetric compactification, the five-branes have to wrap a
holomorphic curve in the Calabi-Yau space $X$, whose second homology
class we denote by $W\in H_2(X,{\mathbb Z})$.

\vskip 0.4cm

Two additional conditions need to be imposed on this data if the
associated compactification is to preserve $N=1$ supersymmetry in four
dimensions. First, the two bundles $V$ and $\tilde{V}$ need to be
(semi-) stable bundles \cite{duy}. 
To introduce the notion of stability, we define
the slope
\begin{equation}
 \mu (F)=\frac{1}{{\rm rk}(F)}\int_Xc_1(F)\wedge J\wedge J
\end{equation}
of a (coherent) sheaf $F$ on $X$, where $J$ is the K\"ahler form on $X$
and ${\rm rk}(F)$ and $c_1(F)$ are the rank and the first Chern class
of the sheaf, respectively. A bundle $V$ is now called stable (resp.~semi-stable)
if for all sub-sheafs $F\subset V$ with $0<{\rm rk}(F)<{\rm rk}(V)$ the
slope satisfies $\mu (F)<\mu (V)$ (resp.~$\mu (F)\leq \mu (V)$). It is worth
mentioning that a bundle $V$ is semi-stable exactly if its dual $V^*$
is and that $h^0(X,V)=h^3(X,V)=0$ for a stable bundle $V$. To preserve
supersymmetry, semi-stability of the bundle is sufficient,
although in practice one often requires stability. For specific examples,
either condition is typically very hard to check and the stability proof
for the bundles considered in this paper, is one of our main results. 
In addition, for supersymmetry to be preserved, the five-brane class $W$
needs to be an effective class. This means that there indeed exists
a holomorphic curve with class $W$ in $X$. 

Finally, heterotic models need to satisfy a well-known anomaly condition.
For the case of bundles $V$ and $\tilde{V}$ with vanishing first Chern classes,
$c_1(V)=c_1(\tilde{V})=0$, which we consider in this paper this condition reads
\begin{equation}
 c_2(TX)-c_2(V)-c_2(\tilde{V})=W\; . \label{anomaly}
\end{equation}

\vskip 0.4cm

Next, we turn to the general structure of the low-energy
particle spectrum. In addition to the dilaton, $h^{1,1}(X)$ K\"ahler
moduli and $h^{2,1}(X)$ complex structure moduli of the Calabi-Yau
space, each of the $E_8$ gauge theories as well as the five-branes
give rise to a sector of particles in the low-energy theory. Here, we
will focus on the ``observable'' sector, associated to the first $E_8$
gauge theory with vector bundle $V$ and structure group $G$. We will not
explicitly consider the particle content in the other ``hidden''
sectors.

The low-energy gauge group $H$ in the observable sector is given by
the commutant of the structure group $G$ within $E_8$. For
$G={\rm SU}(3)$, ${\rm SU}(4)$, ${\rm SU}(5)$ this implies the standard
grand unified groups $H=E_6$, ${\rm SO}(10)$, ${\rm SU}(5)$, respectively.
In order to find the matter field representations, we have to
decompose the adjoint ${\bf 248}$ of $E_8$ under $G\times H$. In
general, this decomposition can be written as
\begin{equation}
248\rightarrow (1,\mbox{Ad}(H))\oplus \bigoplus_{i}(R_{i},r_{i})  
\label{adjoint}
\end{equation}
where $\mbox{Ad}(H)$ denotes the adjoint representation of $H$ and
$\{(R_{i},r_{i})\}$ is a set of representations of $G\times H$.
The adjoint representation of $H$ corresponds to the low-energy
gauge fields while the low-energy matter fields transform in the representations
$r_i$ of $H$. For the three relevant structure groups these matter
field representations are explicitly listed in Table 1.
\begin{table}
\begin{center}
\begin{tabular}{|l|l|}
\hline
$E_{8}\rightarrow G\times H$ & Residual Group Structure \\  \hline\hline
$\rm{SU}(3)\times E_{6}$ & ${\bf 248}\rightarrow ({\bf 1},{\bf 78})\oplus ({\bf 3},{\bf 27})\oplus
(\overline{\bf 3},\overline{\bf 27})\oplus ({\bf 8},{\bf 1})$ \\  \hline
$\rm{SU}(4)\times\rm{SO}(10)$ &$ {\bf 248}\rightarrow ({\bf 1},{\bf 45})\oplus ({\bf 4},{\bf 16})
\oplus (\overline{\bf 4},\overline{\bf 16})\oplus ({\bf 6},{\bf 10})\oplus ({\bf 15},{\bf 1})$ \\  \hline
$\rm{SU}(5)\times\rm{SU}(5)$ & ${\bf 248}\rightarrow ({\bf 1},{\bf 24})\oplus
({\bf 5},\overline{\bf 10})\oplus (\overline{\bf 5},{\bf 10})\oplus ({\bf 10},{\bf 5})\oplus
(\overline{\bf 10},\overline{\bf 5})\oplus ({\bf 24},{\bf 1})$
 \\ \hline
\end{tabular}
\label{rep}
\caption{\em Breaking patterns of $E_8$ and decompositions of
  the ${\bf 248}$ adjoint representation.}
\end{center}
\end{table}
We may ask how many supermultiplets will occur in the low energy theory for
each representation $r_i$? It turns out that this number is given by
$n_{r_{i}}=h^{1}(X,V_{R_{i}})$, the dimension of the cohomology group 
$H^{1}(X,V_{R_{i}})$
of the vector bundle $V$ in the specific $G$ representation $R_i$ which is
paired up with the $H$ representation $r_i$ in the decomposition~\eqref{adjoint}.
For $G={\rm SU}(n)$, the relevant representations $R_i$ can be obtained
by appropriate tensor products of the fundamental representation and one ends
up having to compute  $h^{1}(X,V\otimes V^{* })$, $%
h^{1}(X,V)$, $h^{1}(X,V^{* })$, $h^{1}(X,\wedge ^{2}V)$, and $%
h^{1}(X,\wedge ^{2}V^{* })$. Using Serre duality, $h^{1}(X,V^{*
})=h^{2}(X,V)$, 
the number the low-energy representations can then be computed as
summarized in Table 2.
\begin{table}
\begin{center}
\begin{tabular}{|l|l|} \hline
Decomposition & Cohomologies \\ \hline
$\rm{SU}(3)\times E_{6}$ & $n_{27}=h^{1}(V), n_{\overline{27}}=h^{1}(V^{*
})=h^{2}(V), n_{1}=h^{1}(V\otimes V^{* })$ \\ \hline 
$\rm{SU}(4)\times \rm{SO}(10)$ &$ n_{16}=h^{1}(V), n_{\overline{16}}=h^{2}(V), 
n_{10}=h^{1}(\wedge ^{2}V), n_{1}=h^{1}(V\otimes V^{* })$ \\ \hline 
$\rm{SU}(5)\times\rm{SU}(5)$ & $n_{10}=h^{1}(V^*), n_{\overline{10}}=h^{1}(V), 
n_{5}=h^{1}(\wedge^{2}V), n_{\overline{5}}=h^{1}(\wedge ^{2}V^{* })$\\
&$n_{1}=h^{1}(V\otimes V^{* })$ \\ \hline
\end{tabular}
\label{spec}
\caption{\em Computation of low-energy particle spectra.}
\end{center}
\end{table}
Further, the Atiyah-Singer index theorem \cite{hart}, applied to the case
$c_1(TX)=c_1(V)=0$, tells us that the index of $V$ can be expressed as 
\begin{equation}
\ind(V)=\sum_{p=0}^3 (-1)^{p}\, h^{p}(X,V)=\frac{1}{2}\int_{X}c_3(V) \ ,
\label{index}
\end{equation}
where $c_3(V)$ is the third Chern class of $V$. For a stable bundle,
we have $h^0(X,V)=h^3(X,V)=0$ and comparison with Table 2 shows
that, in this case, the index counts the chiral asymmetry, that is,
the difference of
the number of generations and anti-generations. The index is usually easier
to compute than individual cohomologies and is useful
to impose a physical constraint on the chiral asymmetry.

\vskip 0.4cm

The heterotic models considered in this paper will be constructed
as follows. After choosing a Calabi-Yau space $X$
(which we will take to be one of the five Calabi-Yau spaces realized
as intersections in a single ordinary projective space), we will scan
over a certain, well-defined class of (monad) bundles, $V$, on $X$.  We
will think of these bundles as bundles in the observable sector and
take the hidden bundle $\tilde{V}$ to be trivial. The anomaly
condition~\eqref{anomaly} can then be satisfied by including
five-branes as long as $c_2(TX)-c_2(V)$ is an effective class on $X$.
This is precisely what we will require. In addition, we will only
consider bundles $V$ whose index is a (non-zero) multiple of three.
Only such bundles have a chance, after dividing out by a discrete symmetry,
of producing a model with chiral asymmetry three. We will then prove
stability for all such bundles and compute their complete low-energy spectrum.

%
%==========
%
\section{Monad Construction of Vector Bundles}\label{s:monad}
To begin our systematic construction of vector bundles for heterotic
compactifications, we will make use of a standard and powerful
technique for defining bundles, known as the {\it monad
  construction}. On complex projective varieties, this method of
constructing vector bundles dates back to the early works on $\IP^4$
by \cite{HM} and systematic approaches by \cite{beilinson} and
\cite{maruyama}. This construction defines a vast class of vector
bundles; in fact, every bundle on $\IP^n$ can be expressed as a monad
\cite{barth,HM}. Bundles defined as monads have been widely used in
the mathematics and physics literature. The reader is referred to
\cite{monadbook} for the most general construction of monads and their
properties.  In this work we will use a restricted form prevalent in
the physics literature.
\subsection{The Calabi-Yau Spaces}
Our monad bundles will be constructed on complete intersection
Calabi-Yau manifolds, $X$, which are defined in a single projective ambient space
${\cal A}=\IP^m$. There are five such Calabi-Yau manifolds~\cite{hubsch}
and their properties are summarized in Table 3.
\begin{table}
\begin{center}
\begin{tabular}{|c|c|c|c|c|c|c|c|}\hline
Intersection&${\cal A}$ &Configuration& $\chi (X)$ & $h^{1,1}(X)$ &
 $h^{2,1}(X)$ & $d(X)$ & $\tilde{c}_2(TX)$ \\ \hline
Quintic &$\IP^4$& $[4|5]$ & $-200$ & $1$ & $101$&$5$ & $10$ \\
Quadric and quartic& $\IP^5$ & $[5|2 \ 4]$ & $-176$ & $1$ & $89$ &$8$&
  $7$ \\
Two cubics&$\IP^5$ & $[5|3 \ 3]$ &$-144$ & $1$ &$73$ &$9$& $6$ \\
Cubic and 2 quadrics&$\IP^6$ & $[6|3 \ 2 \ 2]$ & $-144$ & $1$ &
  $73$ & $12$ & $5$ \\
Four quadrics&$\IP^7$ & $[7|2 \ 2 \ 2 \ 2]$ & $-128$ & $1$ & $65$ & $16$&
  $4$ \\ \hline
\end{tabular}
\caption{\em The five complete intersection Calabi-Yau manifolds in a single
projective space. Here, $\chi (X)$ is the Euler number, $h^{1,1}(X)$ and $h^{2,1}(X)$
are the Hodge numbers, $d(X)$ is the intersection number and $c_2(TX)=\tilde{c}_2(TX)J^2$
is the second Chern class. The normalization of the K\"ahler form $J$ is defined in
the main text.\label{t:cy}} 
\end{center}
\end{table}
They are most conveniently described by the configurations
$[m|q_1,\ldots ,q_K]$ listed in the Table, where $m$ refers to the
dimension of the ambient space $\mathbb{P}^m$ and the numbers $q_a$
indicate the degree of the defining polynomials. In this notation the
Calabi-Yau condition $c_1(TX)=0$ translates to $\sum_{a=1}^Kq_a=m+1$.
Furthermore, note that $h^{1,1}(X)=1$ for all five
cases. Hence, these manifolds have their Picard group,
${\rm Pic}(X)$, being isomorphic to ${\mathbb Z}$. 
Such manifolds are called {\it cyclic} \cite{jardim}. 
The K\"ahler form $J$
descends from the the ambient space $\IP^n$ and is normalized
as
\begin{equation}
 \int_{\IP^n}J^m=1\; .
\end{equation}
Integrals over $X$ of any three-form $w$, defined on ${\cal A}=\IP^m$,
can be reduced to integrals over the ambient space using the formula
\begin{equation}
 \int_Xw=d(X)\int_{\IP^m}w\wedge J^{m-3}\; ,
\end{equation}
where $d(X)$ are the intersection numbers listed in Table 3.
The second homology $H_2(X,\mathbb{Z})$ is dual to the integer multiples of
$J\wedge J$ and the Mori cone of $X$ corresponds to all positive multiples
of $J\wedge J$~\cite{Hosono:1994ax}.

For our subsequent analysis it is useful to discuss some properties of line
bundles on the above Calabi-Yau manifolds. We denote by $\cO (k)$ the $k^{\rm th}$
power of the hyperplane bundle, $\cO(1)$, on the ambient space $\IP^m$ and by
$\cO_X (k)$ its restriction to the Calabi-Yau space $X$. The normal bundle $\cN$
of $X$ in the ambient space is then given by
\beq
 \cN=\bigoplus_{a=1}^K\cO (q_a)\; . \label{normal}
\eeq
In general, one finds, for the Chern characters of line bundles on $X$,
\bea
{\rm ch}_1(\cO_X(k))&=&c_1(\cO_X(k))=kJ \ ,\\
{\rm ch}_2(\cO_X(k))&=&\frac{1}{2}k^2J^2 \ ,\\
{\rm ch}_3(\cO_X(k))&=&\frac{1}{6}k^3J^3\; .
\eea
From the Atiyah-Singer index theorem the index of $\cO_X(k)$ is given by
\bea
 {\rm ind}(\cO_X(k))&\equiv&\sum_{q=0}^3(-1)^qh^q(X,\cO_X(k))\nn \\
  &=&\int_X\left[{\rm ch}_3(\cO_X(k))+\frac{1}{12}c_2(TX)\wedge
   c_1(\cO_X(k))\right]\nn \\ 
  &=&\frac{d(X)k}{6}\left(k^2+\frac{1}{2}\tilde{c}_2(TX)\right)\; , 
\label{indline}
\eea
where the numbers $\tilde{c}_2(TX)$ characterize the second Chern class of $X$
and $d(X)$ are the intersection numbers. The values for these quantities
can be read off from Table 3.

We recall that the Kodaira vanishing theorem \cite{hart} states that
on a K\"ahler manifold $X$, $H^q(X, L \otimes K_X)$ vanishes for $q>0$
and $L$ a positive line bundle. Here, $K_X$ is the canonical bundle on
$X$. For Calabi-Yau manifolds $K_X$ is of course trivial and, hence,
the only non-vanishing cohomology for positive line
bundles on Calabi-Yau manifolds is $H^0$. The dimension of this
cohomology group can then be computed from the index theorem.
In fact, inserting the values for the intersection numbers and
the second Chern class from Table 3 into Eq.~\eqref{indline} we 
explicitly find, for the five Calabi-Yau spaces and for line bundles
$\cO_X(k)$ with $k>0$, that
\bea
 h^0([4|5],\cO_X(k))&=&\frac{5}{6}(k^3+5k) \ , \label{cy1}\\
 h^0([5|2\,4],\cO_X(k))&=&\frac{2}{3}(2k^3+7k) \ , \\
 h^0([5|3\,3],\cO_X(k))&=&\frac{3}{2}(k^3+3k) \ , \\
 h^0([6|3\,2\,2],\cO_X(k))&=&2k^3+5k \ , \\
 h^0([7|2\,2\,2\,2],\cO_X(k))&=&\frac{8}{3}(k^3+2k)\; .\label{cy5}
\eea
For negative line bundles $L=\cO_X (-k)$, where $k>0$, it follows
from Serre duality on the Calabi-Yau three-fold $X$, 
$h^q(X,L)=h^{3-q}(X,L^*)$, that only
$H^3(L,X)$ can be non-zero and that its dimension
$h^3(X,\cO_X(-k))=h^0(X,\cO_X(k))$ is given by one of the explicit
expressions~\eqref{cy1}--\eqref{cy5}. Finally, we have
\beq
h^0(X,\cO_X)=h^3(X,\cO_X)=1\; ,\qquad h^1(X,\cO_X)=h^2(X,\cO_X)=0\; .
\label{h0}
\eeq
Now we explicitly know the cohomology for all line bundles on
the five Calabi-Yau manifolds under consideration.
In particular, we conclude that $h^0(X,\cO_X(k))>0$ precisely for $k\geq
0$ and, hence, that only the line bundles $\cO_X(k)$ with $k\geq 0$
have a non-trivial section. This is one of the underlying conditions
for the validity of Hoppe's criterion which will play a central role
in the stability proof for our bundles.

\subsection{Constructing the Monad} 
Having discussed the manifold $X$ and line bundles thereon, we now
construct the requisite vector bundles $V$.
Our construction proceeds as follows. On a Calabi-Yau manifold $X$, a
monad bundle $V$ 
is defined by the short exact sequence 
\beq\label{monad}
0 \to V \stackrel{f}{\longrightarrow} B \stackrel{g}{\longrightarrow}
C \to 0 \; ,
\eeq
where $B$ and $C$ are bundles on $X$. It is standard to take
$B$ and $C$ to be direct sums of line bundles over $X$, that is
\beq
 B=\bigoplus\limits_{i=1}^{r_B} \cO_X(b_i)\; ,\qquad
 C=\bigoplus\limits_{i=1}^{r_C} \cO_X(c_i)\; .
\eeq
Here, $r_B$ and $r_C$ are the ranks of the bundles $B$ and $C$, respectively.
The exactness of \eref{monad} implies that $\ker(g) = \im(f)$ and $\ker(f)=0$,
so that the bundle $V$ can be expressed as
\[
V = \ker(g) \ .
\]
The map $g$ is a morphism between bundles and can be defined as a $r_C
\times r_B$ matrix whose entries, $(i,j)$, are sections of $\cO_X (c_i-b_j)$.
As we have seen in the previous subsection, such sections exist iff
$c_i\geq b_j$ and so this is what we should require. In fact,
if $c_i=b_j$ for an index pair $(i,j)$ the two corresponding line bundles
can simply be dropped from $B$ and $C$ without changing the resulting bundle $V$.
In the following, we will, therefore, assume the stronger condition $c_i>b_j$
for all $i$ and $j$. 

The Calabi-Yau manifolds discussed in this paper are complete
intersections in a single projective space $\IP^m$. We can, therefore, write down
an analogous short exact sequence
\beq\label{monada}
0 \to \cV \stackrel{\tilde{f}}{\longrightarrow}\cB \stackrel{\tilde{g}}
{\longrightarrow} \cC \to 0 \; ,
\eeq 
on the ambient space where
\beq
 \cB=\bigoplus\limits_{i=1}^{r_B} \cO(b_i)\; ,\qquad
 \cC=\bigoplus\limits_{i=1}^{r_C} \cO(c_i)\; .
\eeq
The map $\tilde{g}$ can be viewed as a  $r_C \times r_B$ matrix
whose entries, $(i,j)$, are homogeneous polynomials of degree $c_i-b_j$.
This sequence defines a vector bundle $\cV$ on the ambient space whose
restriction to $X$ is $V$. Further, the map $g$ can be seen as the
restriction of its ambient space counterpart $\tilde{g}$ to $X$.
Unless explicitly stated otherwise, we will assume throughout that this map is generic.

It is natural to enquire whether $V$ thus defined is always a bona fide
bundle rather than a sheaf. We are assured on this point by the
following theorem \cite{fulton2}.
\begin{theorem}
Over any smooth variety $X$, if $g : B \to C$ is a morphism between 
locally free sheaves $B$ and $C$, then $\ker(g)$ is locally free.
\end{theorem}
Now, by definition, a locally free sheaf of constant rank is a vector
bundle. Therefore, by the above theorem, it only remains to check
whether $\ker(g)$ has constant rank on $X$. Indeed, $g$ could be
less than maximal rank on a singular (sometimes called `degeneracy')
locus. We note that exactness of the sequence, that is ${\rm coker}(g)=0$,
is equivalent to this degeneracy locus being empty.

To show that the degeneracy locus is empty for our bundles, it turns
out to be convenient to consider the dual bundle $V^*$ defined by the
dual sequence
\beq\label{monaddual}
0 \to C^* \stackrel{g^T}{\longrightarrow} B^* 
\longrightarrow V^* \to 0 \; ,
\eeq
where
\beq\label{defV^*}
V^* = \coker(g^T) \ .
\eeq
We can now apply the following theorem~\cite{maria,lazarsfeld}.
\begin{theorem}\label{dualtheorem}
Let $\phi: E \to F$ be a morphism of vector bundles on a
variety of dimension $N$ and let $e = \rk(E)$, $f = \rk(F)$ and $e \le
f$. If $E^* \otimes F$ is globally generated and $f - e + 1 > N$, then
generic maps $\phi$ have a vanishing degeneracy locus.
\end{theorem}
Therefore, take $\phi = g^T$, $E = C^*$ and $F = B^*$.  For all our
bundles of interest, $N=3$ and $e < f$. In fact, $f-e$ is the rank of
$V$, which is 3, 4, or 5 for the bundles of interest in heterotic
compactifications. Finally, $E^* \otimes F$ is globally generated
because $B$ and $C$ are direct sums of line bundles with $c_i > b_j$
for all $i,j$. Hence, all the conditions in the theorem are obeyed and
we see that the degeneracy locus of $g^T$, and hence the one for $g$,
is vanishing for the bundles of interest on the Calabi-Yau. However,
one should note that this criterion will not always be satisfied when
writing monad sequences on the higher dimensional ambient spaces, as in
Eq.~\eqref{monada}. (Such
issues will be discussed further in section 4.4). For more on the
degeneracy locus of bundle maps, and why Theorem \ref{dualtheorem}
guarantees its vanishing in the dual monad, see e.g.~\cite{munoz,laytimi}.)

For later reference we present the formulae for the Chern classes of
$V$ (see Ref.~\cite{hubsch}). Simplifying the expressions for $c_2(V)$ and
$c_3(V)$ by imposing the vanishing of the first Chern class, we have
\bea
 \rk(V) &=& r_B - r_C \ , \\ 
 c_1(V) &=& \left(\sum\limits_{i=1}^{r_B} b_i - \sum\limits_{i=1}^{r_C}c_i\right)
  J \equiv 0 \ , \label{c1} \\
 c_2(V) &=& -\frac12 \left(\sum\limits_{i=1}^{r_B} b_i^2 - 
  \sum\limits_{i=1}^{r_C} c_i^2\right) J^2 \ , \label{c2}\\
 c_3(V) &=& \frac13 \left(\sum\limits_{i=1}^{r_B} b_i^3 - 
  \sum\limits_{i=1}^{r_C} c_i^3\right) J^3 \ .
\eea
Hence, from Eq.~\eqref{index} and the above expression for the third Chern class,
the index of $V$ is explicitly given by
\begin{equation}
 {\rm ind}(V)= \sum_{p=0}^3(-1)^p\,h^p(X,V)=
 \frac{d(X)}{6}\left(\sum\limits_{i=1}^{r_B} b_i^3 - 
  \sum\limits_{i=1}^{r_C} c_i^3\right)\; . \label{indV}
\end{equation}
Within this paper, we will make extensive use of the computer algebra system
\cite{m2} in analyzing the monads in \eref{monad}. Utilizing this
powerful tool we are able to catalog efficiently bundle cohomologies
previously too difficult to be calculated. Indeed, computing particle
spectra, that is, 
sheaf cohomology, is ordinarily a tremendous task even for a single bundle,
and it would be unthinkable to attempt to calculate by hand the
hundreds of such cohomologies necessary in a systematic study of
monad bundles. However, the recent advances in algorithmic
algebraic geometry allow us to explicitly and efficiently compute the
requisite cohomology groups for a certain class of bundles.
For the first time, we describe in detail how to use this technology
in the context of string compactification.

With this approach in mind, we recall that in computational algebraic
geometry \cite{schenck}, sheafs are expressed in the language of
graded modules over polynomial rings. If $X$ is embedded in $\IP^m$
with homogeneous coordinates $[x_0:x_1:\ldots:x_m]$, we can let $R$ be
the coordinate ring $\IC[x_0, x_1, \ldots, x_m] / (X)$ where $(X)$ is
the ideal associated with $X$. The bundles $B$ and $C$ are then
described by free-modules of $R$ with appropriate degrees
(grading). We leave to the Appendix a detailed tutorial of the
sheaf-module correspondence and the construction and relevant
computation of monad bundles using computer algebra.

\subsection{Stability of Monad Bundles}
\label{sec:stability}
As mentioned in the previous section, (semi-)stability of the vector
bundle is of central
importance to heterotic compactifications. In general, proving
stability is an overwhelming technical obstacle and a systematic analysis
has so far been elusive.
However, for a class of manifolds, a sufficient but by no means
necessary condition is of great utility; this is the so-called Hoppe's
criterion \cite{hoppe,huybrechts}:
\begin{theorem}\label{hoppe}
[Hoppe's Criterion]
Over a projective manifold $X$ with Picard group $Pic(X) \simeq \IZ$
(i.e., $X$ is cyclic),
let $V$ be a vector bundle with $c_1(V)=0$. If $H^0(X, \bigwedge^p V) = 0$
for all $p = 1, 2, \ldots, \rk(V) - 1$, then $V$ is stable.
\end{theorem}
We also recall that for the Calabi-Yau manifolds used in this paper all
positive line bundles have a section, an underlying assumption for the
validity of Hoppe's theorem which is, hence, satisfied.

The strategy is therefore clear. To prove stability for the monad
bundles \eref{monad} over cyclic manifolds $X$ using
Hoppe's criterion, we need to show the vanishing of $H^0(X, \wedge^p
V)$ for $p=1,\ldots,\rk(V)-1$. In the following paragraphs, we will
outline the basis for this stability proof and make note of certain
results and properties that are of particular use.

One additional assumption which we will make is that all line bundles
involved in the definition of the bundles $V$ are positive, that is,
for all $i$,
\beq\label{positive}
b_i>0 \quad \mbox{and }c_i>0 \; .
\eeq
We will refer to this property as ``positivity'' of the bundle $V$.
While this is not required for a consistent definition of the bundle
or the associated heterotic model, it turns out to be
a crucial technical assumption which facilitates the stability
proof. The essential point is that  positivity of $V$ allows one to use Kodaira
vanishing when applying Hoppe's criterion to the dual bundle $V^*$.
To see how this works, recall that the dual bundle is defined by
the sequence  $0 \to C^* \longrightarrow B^* \longrightarrow V^* \to 0$
and that its stability is equivalent to that of $V$. The associated
long exact sequence in cohomology is
\bea
0&\to& H^0(X, C^*) \to H^0(X, B^*) \to \fbox{\mbox{$H^0(X, V^*)$}}\nn\\
 &\to& H^1(X, C^*) \to H^1(X, B^*)\to H^1(X,V^*)\nn\\
 &\to&H^2(X,C^* )\to  H^2(X, B^*)\to H^2(X,V^*)\nn\\
 &\to& H^3(X,C^*) \to H^3(X, B^*) \to H^3(X, V^*) \to 0 \ .
\label{dual-seq}
\eea
Given that we are dealing with positive bundles $V$, it follows that
$B^*$ and $C^*$ are sums of negative line bundles and, hence,
$H^0(X, B^*)$ and
$H^1(X, C^*)$ in the above sequence are zero due to Kodaira vanishing.
This means the ``boxed'' cohomology $H^0(X, V^*)$ also vanishes.
(For later considerations we note that Kodaira vanishing also
implies $H^1(X,B^*)=H^2(X,C^*)=0$ and, hence, $H^1(X,V^* )\simeq H^2(X,V)=0$.)
In order to prove stability of $V^*$ by applying Hoppe's criterion we
have to show that $H^0(X, \wedge^p V^*)=0$ for $p=1,\ldots, \rk(V)-1$
and we have just completed the first step for $p=1$.

Next, we need to compute the cohomologies $H^0X, \wedge^p V^*)$ for $p>1$.
However, a further simplification occurs because we are dealing with
unitary bundles. In fact, for an $SU(n)$ bundle $V$, we have
\beq
\wedge^{n-1} V^* \simeq V \label{wedge}
\eeq
(see, for example Ref.~\cite{fulton}) .
Therefore, to cover the case $p=n-1$, the highest exterior power
relevant to Hoppe's criterion, we only need to show that $H^0(X,V) =
0$.  This is indeed the case for all bundles considered in this paper
and the explicit proof, which is somewhat lengthy, is presented in
Appendix \ref{ap:h0}. This completes the stability proof for the
rank $3$ bundles. 

For rank $4$ and $5$ bundles we have to look at further exterior powers of
$V^*$, namely $\Lambda^pV^*$ for $p = 2, \ldots, \rk(V)-2$. To deal
with those we consider the standard long exact (``exterior power'') 
sequence~\cite{maria,hart} for $\Lambda^pV^*$
\bea
0 &\to& S^p C^* \to S^{p-1} C^* \otimes B^* \to S^{p-2} C^* \otimes
\wedge^2 B^* \to \ldots\nn\\
& \to& A \otimes \wedge^{p-1} B^* \to \wedge^p B^* \to \wedge^p V^* \to 0 \ ,
\label{wedge-seq1}
\eea
which is induced by the short exact sequence \eqref{monaddual}.
Here $S^i$ is the $i$-th symmetrised tensor power of a bundle.
Such a sequence does not itself induce a long exact sequence in
cohomology; we need to slice it up into groups of three. In other
words, we introduce co-kernels $K_i$ such that \eref{wedge-seq1} becomes
the following set of short exact sequences
\bea
0 &\to& S^p C^* \to S^{p-1} C^* \otimes B^* \to K_1 \to 0 \ , \nn\\
0& \to& K_1 \to S^{p-2} C^* \otimes \wedge^2 B^* \to K_2 \to 0 \ , \nn \\
&&\qquad\qquad\qquad\qquad\vdots \nn \\
 0 &\to& K_{p-1} \to \wedge^p B^* \to \wedge^p V^* \to 0 \ .
\label{wedge-seq}
\eea
Each of the above now induces a long exact sequence in cohomology in
analogy to \eref{dual-seq}:
\beq\label{wedge-hom}
{\hspace{-1cm}
\ba{rcl}
&&0 \to H^0(X, S^p C^*) \to H^0(X, S^{p-1} C^* \otimes B^*) \to H^0(X,
K_1) \to H^1(X, S^p C^*) \to \ldots \to 0 \ , \\
&&0 \to H^0(X, K_1) \to H^0(X, S^{p-2} C^* \otimes \wedge^2 B^*) 
\to H^0(X, K_2) \to H^1(X, K_1) \to \ldots \to 0 \ , \\
&& \hspace{6cm} \vdots\\
&&0 \to H^0(X,  K_{p-1}) \to H^0(X, \wedge^p B^*) \to 
\fbox{\mbox{$H^0(X, \wedge^p V^*)$}}
\to H^1(X,  K_{p-1}) \to \ldots \to 0 \ . \\
\ea
}
\eeq
The term we need is boxed and we need to trace through the various
sequences, using the readily computed cohomologies of the
symmetric and antisymmetric powers of $B^*$ and $C^*$, to arrive at the
answer. Let us now do this explicitly for the case $p=2$, that is, 
$H^0(X,\Lambda^2V^*)$.
The long exact sequence~\eqref{wedge-seq1} then specializes to
\begin{equation}
 0\to S^2C^*\to C^*\otimes B^*\to\Lambda^2B^*\to\Lambda^2
 V^*\to 0\; ,
\end{equation}
which needs to be broken up into the two short exact sequences
\bea
 &0&\to S^2C^*\to C^*\otimes B^*\to K\to 0\label{short1}\\
&0&\to K\to \Lambda^2B^*\to\Lambda^2 V^*\to 0\; .\label{short2}
\eea
From the first of these we have the long exact sequence
\bea
 0&\to&H^0(X,S^2C^* )\to H^0(X,C^*\otimes B^* )\to H^0(X,K)\nn\\
  &\to&H^1(X,S^2C^* )\to H^1(X,C^*\otimes B^* )\to H^1(X,K)\nn\\
  &\to&H^1(X,S^2C^* )\to\ldots\; .
  \label{seq1}
\eea
Since $B^*$ and $C^*$ are sums negative line bundles, so are their
various tensor products which appear in the above sequences. From Kodaira
vanishing all cohomologies of such bundles vanish except for the
third. 
Applying this
to \eqref{seq1} we immediately deduce that $H^0(X,K)=H^1(X,K)=0$. Using this
information in the long exact sequence
\beq
0\to H^0(X,K)\to H^0(X,\Lambda^2B^* )\to H^0(X,\Lambda^2V^* )
\to H^1(X,K)\to\ldots
\eeq
which follows from \eqref{short2} we find $H^0(X,\Lambda^2V^* )=0$,
as desired. This completes the stability proof for rank $4$
bundles~\footnote{Together with $H^0(X,V^* )=0$, which we have shown earlier, it
also provides an independent argument for the stability of rank $3$ bundles.}.

Finally, for rank $5$ bundles, we still need to compute $H^0(X,\Lambda^3V^*)$.
Repeating the above steps for this case one finds that Kodaira vanishing on $X$
alone does not quite provide sufficient information to conclude that
$H^0(X,\Lambda^3V^*)=0$. In this case, we need to employ the additional
technique of  \textit{Koszul sequences} \cite{hubsch,hart} which rely
on the embedding of the Calabi-Yau manifold in an ambient space ${\cal A}$.
Specifically, for a vector bundle $\cW$ on ${\cal A}$ the Koszul sequence reads
\begin{equation}\label{koszul}
0\rightarrow \wedge ^{K}\cN^{* }\otimes \cW\rightarrow ...\rightarrow \wedge
^{2}\cN^{* }\otimes \cW\rightarrow \cN^{* }\otimes \cW\rightarrow \cW\overset{
\rho }{\rightarrow }\cW|_{X}\rightarrow 0 \ ,
\end{equation}
where $\cW|_X$ denotes the restriction of $\cW$ to $X$ and $\rho$ is the
associated restriction map. Here $\cN^*$ is the dual of the Calabi-Yau
normal bundle, defined in Eq.~\eqref{normal}. As will be shown in the
next section, the Koszul sequence can be used to compute the relevant
cohomologies directly from the ambient space. This will allow us to
complete the stability proof for rank 5 bundles.

\section{Classification and Examples}\label{s:eg}
Armed with the general information about the five Calabi-Yau manifolds
and monad bundles we can now proceed to classify such bundles, prove their
stability and compute their spectrum.
\subsection{Classification of Configurations}\label{s:class}
For the monad bundles defined by the short exact
sequence~\eqref{monad}, we can immediately formulate a classification
scheme. Recall that, taking the bundles $B$ and $C$ to be direct sums
of line-bundles over the manifold $X$, we have
\beq\label{monad2}
0 \to V \to \bigoplus\limits_{i=1}^{r_B} \cO_X(b_i) 
\stackrel{g}{\longrightarrow}
\bigoplus\limits_{i=1}^{r_C} \cO_X(c_i) \to 0 \ ,
\qquad
V \simeq \ker(g) \ .
\eeq
From our discussion so far these bundles are subject to a number
of physical and mathematical constraints which can be summarised
as follows:
\begin{enumerate}
\item As discussed earlier we require all $b_i$ and $c_i$
  to be positive; this is a technical assumption which will
  significantly simplify our
  computations.
\item We furthermore require that $b_i < c_j$ for all $i$ and $j$;
  this is to ensure that the map $g$, which consists of sections of
  $\cO_X(c_j-b_i)$, has no zero entries. Further, we require the map
  $g$ to be generic. Then, all conditions of Theorem~\eqref{dualtheorem}
  are met and we are guaranteed that $V$, as defined
  by the sequence~\eqref{monad2}, is indeed a bundle.
\item Since we are dealing with special unitary bundles we impose $c_1(V) = 0$.
\item For a given Calabi-Yau space $X$ and a bundle $V$ we need to ensure
      that the anomaly condition~\eqref{anomaly} can be satisfied. 
      To do this we impose the condition that $c_2(TX) - c_2(V)$ must be effective.
      Then, we can choose a trivial hidden bundle $\tilde{V}$
      and a five-brane wrapping a holomorphic curve with
      homology class  $c_2(TX) - c_2(V)$. In practice, this condition simply means that
      the coefficient of $J^2$ in  $c_2(TX) - c_2(V)$ must be non-negative~\footnote{However,
      for a given example there may well be other ways to satisfy the anomaly
      condition which involve a non-trivial hidden bundle $\tilde{V}$.} .
\item We require that the index of $V$ is a non-zero multiple of three.
      Only such models may lead to three generations after dividing by a discrete
      symmetry.
\item Since we are interested in low-energy grand unified groups we consider bundles
      $V$ with structure group ${\rm SU}(n)$, where $n={\rm rk}(V)=3,4,5$.
\end{enumerate}
Therefore, an integer partitioning problem immediately presents
itself to us: find partitions $\{b_i\}_{i=1,\ldots ,r_C+n}$ and
$\{c_j\}_{i=1,\ldots ,r_C}$ of positive integers $b_i>0$, $c_i>0$ satisfying
$b_i<c_j$ for all $i$, $j$ and subject to the condition
$\sum\limits_{i=1}^{r_B} b_i - \sum\limits_{i=1}^{r_C} c_i = 0$ 
for vanishing first Chern class of $V$ (see Eq.~\eqref{c1}). Further,
we demand that the index of $V$, Eq.~\eqref{indV}, is non-zero and divisible
by three and that the coefficient of $J^2$ in $c_2(TX) - c_2(V)$ be non-negative,
in order to ensure the existence of a holomorphic five-brane curve. From
Eq.~\eqref{c2} the last constraint can be explicitly written as
\beq
0 \le -\frac12 (
\sum\limits_{i=1}^{r_C + n} b_i - \sum\limits_{i=1}^{r_C} c_i)
\le \tilde{c}_2(TX) \ , \label{c2cons}
\eeq
where the numbers $\tilde{c}_2(TX)$ for the second Chern class of $X$
are given in Table 3. Since $b_i<c_j$ for all $i$, $j$ it is clear
that this constraint implies an upper bound on $b_i$ and
$c_j$ and, hence, that the number of vector bundles in our class is
finite~\footnote{The constraint~\eqref{c2cons} arises because we require $N=1$
supersymmetry in four dimensions. If we relaxed this condition and allowed for
anti-five branes there would be no immediate bound on the number of vector
bundles. However, in this case, the stability of such non-supersymmetric models
has to be analyzed carefully~\cite{Gray:2007mg}.}.
To derive this bound explicitly we slightly modify an argument from
Appendix B of Ref.~\cite{Douglas:2004yv}. Define the quantity
\beq
S = \sum\limits_{i=1}^{r_C + n} b_i = \sum\limits_{i=1}^{r_C} c_i \ ,
\eeq
and consider the following chain of inequalities
\bean
2\,\tilde{c}_2(TX) \ge
\sum\limits_{i=1}^{r_C} c_i^2 - \sum\limits_{i=1}^{r_C + n}
b_i^2
& \ge & (b_{max}+1)\sum\limits_{i=1}^{r_C} c_i - 
 \sum\limits_{i=1}^{r_C + n} b_i^2 \\
& = & S + \sum\limits_{i=1}^{r_C + n} b_{max} b_i -
 \sum\limits_{i=1}^{r_C + n} b_i^2 \ge S \ .
\eean
From Table 3, $\tilde{c}_2(TX)$ is at most $10$ and, hence, the sum
$S$ cannot exceed $20$, thereby placing an upper bound on our partitioning
problem.

Given the finiteness of the problem, the classification of all
positive monad bundles subject to the above constraints is now
easily computerisable. Given these conditions, we found $37$ bundles on the
five Calabi-Yau manifolds in question, $20$ for rank $3$, $10$ for rank $4$ and
$7$ for rank $5$. 
Had we relaxed the condition that $c_3$ should be
divisible by 3, we would have found 43, 15, 10, 6, and 3 bundles,
respectively on the 5 cyclic manifolds, for a total of 77.
A complete list of all such bundles for 
the five Calabi-Yau manifolds of concern is given in the Tables 4--8.
\begin{table}
\begin{center}
\begin{tabular}{|c|c|c|c|c|} \hline
Rank & $\{b_i\}$ & $\{c_i\}$ & $c_2(V)/J^2$ & $\ind(V)$ \\ \hline
3 & (2, 2, 1, 1, 1) & (4, 3) & 7 & -60 \\
3 & (2, 2, 2, 1, 1) & (5, 3) & 10 & -105\\ 
3 & (3, 2, 1, 1, 1) & (4, 4) & 8 & -75\\ 
3 & (1, 1, 1, 1, 1, 1) & (2, 2, 2) & 3 & -15\\ 
3 & (2, 2, 2, 1, 1, 1) & (3, 3, 3) & 6 & -45\\ 
3 & (3, 3, 3, 1, 1, 1) & (4, 4, 4) & 9 & -90 \\ 
3 & (2, 2, 2, 2, 2, 2, 2, 2) & (4, 3, 3, 3, 3) & 10 & -90\\ 
3 & (2, 2, 2, 2, 2, 2, 2, 2, 2) & (3, 3, 3, 3, 3, 3) & 9 & -75\\ 
4 & (2, 2, 1, 1, 1, 1) & (4, 4) & 10 & -90\\ 
4 & (1, 1, 1, 1, 1, 1, 1) & (3, 2, 2)  & 5 & -30\\ 
4 & (2, 2, 2, 1, 1, 1, 1) & (4, 3, 3) & 9 & -75\\ 
4 & (2, 2, 2, 2, 1, 1, 1, 1) & (3, 3, 3, 3) & 8 & -60\\ 
5 & (1, 1, 1, 1, 1, 1, 1, 1) & (3, 3, 2) & 7 & -45\\ 
5 & (1, 1, 1, 1, 1, 1, 1, 1) & (4, 2, 2) & 8 & -60\\ 
5 & (2, 2, 2, 2, 2, 1, 1, 1, 1, 1) & (3, 3, 3, 3, 3) & 10 & -75\\ \hline
\end{tabular}
\caption{\em Positive monad bundles on the quintic, $[4|5]$.}
\end{center}
\end{table}
\begin{table}
\begin{center}
\begin{tabular}{|c|c|c|c|c|} \hline
Rank & $\{b_i\}$ & $\{c_i\}$ & $c_2(V)/J^2$ & $\ind(V)$ \\ \hline
3& (2, 2, 1, 1, 1) & (4, 3) & 7 & -96\\ 
3& (1, 1, 1, 1, 1, 1) & (2, 2, 2) & 3 & -24\\ 
3& (2, 2, 2, 1, 1, 1) & (3, 3, 3) & 6 & -72\\ 
4& (1, 1, 1, 1, 1, 1, 1) & (3, 2, 2) & 5 & -48\\ 
5& (1, 1, 1, 1, 1, 1, 1, 1) & (3, 3, 2) & 7 & -72 \\ \hline
\end{tabular}
\caption{\em Positive monad bundles on $[5|2 \ 4]$.}
\end{center}
\end{table}
\begin{table}
\begin{center}
\begin{tabular}{|c|c|c|c|c|} \hline
Rank & $\{b_i\}$ &$ \{c_i\}$ & $c_2(V)/J^2$ & $\ind(V)$ \\ \hline
3& (1, 1, 1, 1) & (4) & 6 & -90\\ 
3& (1, 1, 1, 1, 1) & (3, 2) & 4 & -45\\ 
3& (2, 1, 1, 1, 1) & (3, 3) & 5 & -63\\ 
3& (1, 1, 1, 1, 1, 1) & (2, 2, 2) & 3 & -27 \\ 
3& (2, 2, 2, 1, 1, 1) & (3, 3, 3) & 6 & -81\\ 
4& (1, 1, 1, 1, 1, 1) & (3, 3) & 6 & -72\\ 
4& (1, 1, 1, 1, 1, 1, 1) & (3, 2, 2) & 5 & -54\\
4& (1, 1, 1, 1, 1, 1, 1, 1) & (2, 2, 2, 2) & 4 & -36\\ 
5& (1, 1, 1, 1, 1, 1, 1, 1, 1) & (3, 2, 2, 2) & 6 & -63\\ 
5& (1, 1, 1, 1, 1, 1, 1, 1, 1, 1) & (2, 2, 2, 2, 2) & 5 & -45\\ \hline
\end{tabular}
\caption{\em Positive monad bundles on $[5|3 \ 3]$.}
\end{center}
\end{table}
\begin{table}
\begin{center}
\begin{tabular}{|c|c|c|c|c|} \hline
Rank & $\{b_i\}$ & $\{c_i\}$ & $c_2(V)/J^2$ & $\ind(V)$ \\ \hline
3& (1, 1, 1, 1, 1) & (3, 2) & 4 & -60\\ 
3& (2, 1, 1, 1, 1) & (3, 3) & 5 & -84\\ 
3& (1, 1, 1, 1, 1, 1) & (2, 2, 2) & 3 & -36\\ 
4& (1, 1, 1, 1, 1, 1, 1) & (3, 2, 2) & 5 & -72\\ 
4& (1, 1, 1, 1, 1, 1, 1, 1) & (2, 2, 2, 2) & 4 & -48\\ 
5& (1, 1, 1, 1, 1, 1, 1, 1, 1, 1) & (2, 2, 2, 2, 2) & 5 & -60\\
\hline
\end{tabular}
\caption{\em Positive monad bundles on $[6|2 \ 2 \ 3]$.}
\end{center}
\end{table}
\begin{table}
\begin{center}
\begin{tabular}{|c|c|c|c|c|} \hline
Rank & $\{b_i\}$ & $\{c_i\}$ & $c_2(V)/J^2$ & $\ind(V)$ \\ \hline
3 & (1, 1, 1, 1, 1, 1) & (2, 2, 2) & 3 & -48 \\
\hline
\end{tabular}
\caption{\em Positive monad bundles on $[7|2 \ 2 \ 2 \ 2]$.}
\end{center}
\end{table}
\subsection{$E_6$-GUT Theories}
The first case we shall analyse is $E_6$-GUT theories which arise
from $SU(3)$ bundles. We have already seen in Section~\ref{sec:stability}
that all such bundles are indeed stable. This result has been explicitly
confirmed by a computer algebra computation of $H^0(X,V^*)$ and
$H^0(X,\Lambda^2V^*)$ along the lines described in Appendix~\ref{appA}.
We can, therefore, directly turn to a computation of their particle spectrum.
\subsubsection{Particle Content}
The number of ${\bf 27}$ and $\overline{\bf 27}$ representation of
$E_6$ is easy to obtain. Since $V$ is stable we already know that
$H^0(X,V)=H^3(X,V)=0$. From the long exact sequence~\eqref{dual-seq}
we have deduced earlier that $H^2(X,V)\simeq H^1(X,V^*)=0$
so that $H^1(X,V)$ is the only non-vanishing cohomology. Its
dimension can be directly computed from the index~\eqref{indV}, so that
\beq\label{E6-27}
n_{27} = h^1(X,V) = -\ind(V)\; ,\qquad n_{\overline{27}} = h^2(X,V)=0 \; .
\eeq
Therefore, for the rank 3 bundles in Tables 4--8, the (negative of
the) right-most column gives the number of ${\bf 27}$ representations.
This result also provides the first example of what is a general feature
of positive monad bundles, namely the absence of anti-generations.
The numbers $n_{27}$ have been independently verified by computer algebra.

What about the $E_6$ singlets? These correspond to the cohomology
$H^1(X, \text{ad}(V)) = H^1(X, V \otimes V^*)$. We begin by
tensoring the defining sequence \eqref{monaddual} for $V^*$ by $V$.
This leads to a new short exact sequence
\beq
0 \to C^* \otimes V \to B^* \otimes V \to V^* \otimes V \to 0 \ .
\eeq
One can produce two more short exact sequences by multiplying
\eqref{monaddual} with $B$ and $C$. Likewise, three short exact
sequences can be obtained by multiplying the original sequence~\eqref{monad}
for $V$ with $V^*$, $B^*$ and $C^*$. The resulting six
sequences can then be arranged into the following web of three horizontal
sequences $h_{I}$, $h_{II}$, $h_{III}$ and three vertical ones
$v_I$, $v_{II}$, $v_{III}$.
\beq\ba{cccccccccl}
&&0&&0&&0&&& \\
&&\downarrow&&\downarrow&&\downarrow&&& \\
0&\to& C^* \otimes V &\to& B^* \otimes V &\to& V^* \otimes V &\to&0
\qquad &h_{I} \\
&&\downarrow&&\downarrow&&\downarrow&&& \\
0&\to& C^* \otimes B &\to& B^* \otimes B &\to& V^* \otimes B &\to&0
\qquad &h_{II} \\
&&\downarrow&&\downarrow&&\downarrow&&& \\
0&\to& C^* \otimes C &\to& B^* \otimes C &\to& V^* \otimes C &\to&0
\qquad &h_{III}  \\
&&\downarrow&&\downarrow&&\downarrow&&& \\
&&0&&0&&0&&& \\
&&v_I&&v_{II}&&v_{III}&&& \\
\ea
\eeq
The long exact sequence in cohomology induced by $h_{I}$ reads
\bea\label{VVseq}
0 &\to& H^0(X,C^* \otimes V) \to H^0(X,B^* \otimes V)
\to H^0(X, V^* \otimes V)\nn\\
&\to&H^1(X,C^* \otimes V) \to
 H^1(X, B^* \otimes V)\to \fbox{\mbox{$H^1(X,V^* \otimes V)$}}\nn \\
&\to& H^2(X,C^* \otimes V) \to \ldots\label{h1}
\eea
and we have boxed the term which we would like to compute. We will also need the
long exact sequences which follow from $v_{I}$ and $v_{II}$. They
are given by
\bea
 0&\to&H^0(X,C^*\otimes V)\to H^0(X,C^*\otimes B)\to H^0(X,C^*\otimes C)\nn\\ 
  &\to&H^1(X,C^*\otimes V)\to H^1(X,C^*\otimes B)\to H^1(X,C^*\otimes C)\nn\\
 &\to&H^2(X,C^*\otimes V)\to H^2(X,C^*\otimes B)\to H^2(X,C^*\otimes C)
 \to\ldots\label{v1}\\[0.3cm]
 0&\to&H^0(X,B^*\otimes V)\to H^0(X,B^*\otimes B)\to H^0(X,B^*\otimes C)\nn\\ 
  &\to&H^1(X,B^*\otimes V)\to H^1(X,B^*\otimes B)\to H^1(X,B^*\otimes C)\nn\\
 &\to&H^2(X,B^*\otimes V)\to H^2(X,B^*\otimes B)\to H^2(X,B^*\otimes C)\to\ldots
 \label{v2}
 \eea
Now, because of the integers defining $B$ and $C$ satisfy  $b_i < c_j$,
the tensor product $C^* \otimes B$ is a direct
sum of negative line bundles and, hence, all its cohomology groups
vanish except the third. Further, the middle cohomologies
$H^1$ and $H^2$ of $B^*\otimes B$ and $C^*\otimes C$ vanish.
From the sequence~\eqref{v1} this implies
\beq
 H^0(X,C^*\otimes V)=H^2(X,C^*\otimes V)=0\; , \qquad 
 H^1(X,C^*\otimes V)=  H^0(X,C^*\otimes C)\; .
\eeq
Vanishing of $H^2(X,C^*\otimes V)$ means that the long exact sequence
\eqref{h1} breaks after the second line and we get
\beq
 h^1(X,V^*\otimes V)=h^1(X,B^*\otimes V)-h^1(X,C^*\otimes V)
 +h^0(X,V^*\otimes V)
 -h^0(X,B^*\otimes V)\; .
 \label{nsing0}
\eeq
Using the additional information
\beq
 h^1(X,B^*\otimes V)- h^0(X,B^*\otimes V)=h^0(X,B^*\otimes C)
 -h^0(X,B^*\otimes B)\; .\\
\eeq
which follows from the sequence \eqref{v2} and the fact that
$h^0(X,V^*\otimes V)=1$ (see Theorem B.1 of Ref.~\cite{hubsch})
Eq.~\eqref{nsing0} can be re-written as
\beq
 h^1(X,V^*\otimes V)=h^0(X,B^*\otimes C)-h^0(B^*\otimes B)
                        -h^0(C^*\otimes C)+1\; .\label{nsing}
\eeq
This equation, together with Eqs.~\eqref{cy1}--\eqref{cy5} and
\eref{h0}, allows us to directly compute the number $n_1$ of
$E_6$-singlets and the results are given in \tref{t:E6}. For
reference, we have also included the number of ${\bf
  27}$-representations (the number of $\overline{\bf 27}$ particles,
we recall, is zero). In addition, the results for $h^1(X,V^*\otimes
V)$ have been independently confirmed using Macaulay~\cite{m2},
following the procedure outlined in Appendix~\ref{appA}. We note
that the above derivation of Eq.~\eqref{nsing} is independent
of the rank of the vector bundle $V$ and, hence, it remains
valid for rank $4$ and $5$ bundles.
\begin{table}[h]
\begin{center}
\begin{tabular}{|c|c|c|c|c|} \hline
$X$ & $\{b_i\}$ & $\{c_i\}$ & $n_{27}$ &$ n_1$ 
\\ \hline
% [a,b] in comments means [h^1(B^* \otimes V), h^0(C^*\otimes C)]
$[4|5]$ & (2, 2, 1, 1, 1) & (4, 3) & 60 & 141 %[147, 7]
\\
& (2,2,2,1,1) & (5, 3) & 105 & 231 %[247,17]
\\
& (3, 2, 1, 1, 1) & (4, 4) & 75 & 171 %[174,4]
\\
& (1, 1, 1, 1, 1, 1) & (2, 2, 2) & 15 & 46 %[54,9]
\\
& (2, 2, 2, 1, 1, 1) & (3, 3, 3) & 45 &109 %[117,9]
\\
& (3, 3, 3, 1, 1, 1) & (4, 4, 4) & 90 & 199 %[207,9]
\\
& (2, 2, 2, 2, 2, 2, 2, 2) & (4, 3, 3, 3, 3) & 90 & 180 %[216,37]
\\
& (2, 2, 2, 2, 2, 2, 2, 2, 2) & (3, 3, 3, 3, 3, 3)& 75 & 154 %[189,36]
\\
\hline
$[5|2 \ 4]$ & (2, 2, 1, 1, 1) & (4, 3) & 96 & 206 %[213,8]
\\
& (1, 1, 1, 1, 1, 1) & (2, 2, 2) & 24 & 64 %[72,9]
\\
& (2, 2, 2, 1, 1, 1) & (3, 3, 3) & 72 & 154 %[162,9]
\\
\hline
$[5|3 \ 3]$& (1, 1, 1, 1) & (4) & 90 & 200 %[200,1]
\\
& (1, 1, 1, 1, 1) & (3, 2) & 45 & 103 %[110,8]
\\
& (2, 1, 1, 1, 1) & (3, 3) & 63 & 136 %[139,4]
\\
& (1, 1, 1, 1, 1, 1) & (2, 2, 2) & 27 & 64 %[72,9]
\\
& (2, 2, 2, 1, 1, 1) & (3, 3, 3) & 81 & 163 %[171,9]
\\
\hline
$[6|2 \ 2 \ 3]$& (1, 1, 1, 1, 1) & (3, 2) & 60 & 132 %[140,9]
\\
& (2, 1, 1, 1, 1) & (3, 3) & 84 & 174 %[177,4]
\\ 
& (1, 1, 1, 1, 1, 1) & (2, 2, 2) & 36 & 82 %[90,9]
\\
\hline
$[7|2 \ 2 \ 2 \ 2]$ & (1, 1, 1, 1, 1, 1) & (2, 2, 2) &48 & 100 %[108,9]
\\
\hline
\end{tabular}
{\it \caption{The particle content for the $E_6$-GUT theories arising from
  our classification of stable, positive $SU(3)$ monad bundles $V$ on the
  Calabi-Yau threefold $X$. The number $n_{\overline{27}}$ of
  anti-generations vanishes.}
\label{t:E6}
}
\end{center}
\end{table}
\subsection{$SO(10)$-GUT Theories}
Grand Unified theories with gauge group ${\rm SO}(10)$ are obtained
from rank $4$ bundles with structure group ${\rm SU}(4)$. We have
already shown the stability of positive rank $4$ monad bundles $V$ in
Section~\ref{sec:stability}. As before, we have explicitly confirmed this general
result for the rank $4$ bundles in our classification with Macaulay~\cite{m2},
by showing that $H^0(X,\Lambda^pV^*)$ for $p=1,2,3$ vanishes.
We proceed to analyze the particle content of ${\rm SO}(10)$ GUT
theories.
\subsubsection{Particle Content}
Recall from Table 2, that for $SO(10)$-GUT theories we need to
compute $n_{16}=h^{1}(X,V)$, $n_{\overline{16}}=h^1(X,V^*) = h^{2}(X,V)$,
$n_{10}=h^{1}(X, \wedge ^{2}V)$ and $n_{1}=h^{1}(X, V\otimes V^*)$.

Let us begin with the generations and anti-generations in ${\bf 16}$
and $\overline{\bf 16}$. As in the case of rank $3$
bundles, stability implies that $H^0(X,V)=H^3(X,V)=0$ and, further, from
the sequence~\eqref{dual-seq}, also $H^2(X,V)=H^1(X,V^*)$ is zero.
Hence, as before, the number of anti-generations vanishes and the
number of generations can be computed from the index, so that
\beq
n_{16} =  h^1(X,V) = -\ind(V) \; , \qquad n_{\overline{16}} = 0 \ .
\eeq
Thus, for the rank $4$ bundles in Tables 4--8, the (negative) of
the right-most column gives the number of ${\bf 16}$ representations.

Next, we need to compute the Higgs content which is given by
$n_{10}=h^{1}(X, \wedge ^{2}V)$. It can be shown in general that
for generic maps $g:B\to C$ the number of ${\bf 10}$ representations
always vanishes, that is
\beq
 n_{10}=0\; .
\eeq
The proof is somewhat technical and can be found in Appendix~\ref{ap:h0}.
Again, this result can be readily verified using computer algebra.

Finally, we need to compute the number $n_1$ of ${\rm SO}(10)$ singlets
which is easily obtained from Eq.~\eqref{nsing}. The results for
the spectrum from rank $4$ bundles are summarized in \tref{t:SO10}.

A vanishing number, $n_{10}$, of Higgs particles is not desirable
from a particle physics viewpoint. One might, therefore, wonder whether
more specific choices of the map $g$ in \eqref{monad} could produce a
non-zero value for $n_{10}$. This problem has been encountered
in Ref.~\cite{Bouchard:2005ag,Donagi:2004qk,Donagi:2004ia} where the
spectrum of compactification was shown to depend on the region of
moduli space. 
Specifically, it was shown that the spectrum takes
a generic form with possible enhancements in special regions of the
moduli space; this was dubbed the ``jumping phenomenon'' in
\cite{Donagi:2004qk,Donagi:2004ia}.

To see that a similar phenomenon can arise for monad bundles, let is consider the
following $SU(4)$ bundle on the quintic, $[4|5]$.
\beq\label{5eg-1}
0 \to V \to \cO_X^{\oplus 2}(2) \oplus  \cO_X^{\oplus 4}(1)
\stackrel{g}{\longrightarrow} \cO_X^{\oplus 2}(4) \to 0 \; .
\eeq
This bundle and its particle content for a generic map $g$ is given in
the first line of \tref{t:SO10}. Now we explicitly define the map
$g$ by
\beq\label{5eg-2}
g = \left(
\begin{matrix}
4x_{3}^2& 9x_{0}^2 + x_{2}^2& 8x_{2}^3& 2x_{3}^3&
4x_{1}^3& 9x_{1}^3 \cr
x_{0}^2 + 10x_{2}^2& x_{1}^2& 9x_{2}^3&
7x_{3}^3& 9x_{1}^3 + x_{2}^3& x_{1}^3 + 7x_{4}^3
\end{matrix}
\right) \ .
\eeq
where $x_0,\ldots ,x_4$ are the homogeneous coordinates of $\IP^4$.
This choice for $g$ is no longer completely generic,
although the sequence~\eqref{5eg-1} is still exact.
Following the steps in Appendix \ref{ap:eg}, we can use Macaulay
to calculate the spectrum for this case. We find
\beq
n_{16} =  90\; , \qquad n_{\overline{16}}=0\; ,\qquad
n_{10} = 13\; ,\qquad n_1=277 \; .
\eeq
This is identical to the generic result in Table~\ref{t:SO10},
except for the number of ${\bf 10}$ representations which has
changed from $0$ to $13$.
\begin{table}
\begin{center}
\begin{tabular}{|c|c|c|c|c|c|} \hline
$X$ & $\{b_i\}$ & $\{c_i\}$ & $n_{16}$& $n_1$\\ \hline
%again [a,b] in comments means [h^1(B^* \otimes V), h^0(C^*\otimes C)]
$[4|5]$ & (2, 2, 1, 1, 1, 1) & (4, 4) & 90 & 277 %[280,4]
\\
&  (1, 1, 1, 1, 1, 1, 1) & (3, 2, 2) & 30  & 112 %[126,15]
\\
& (2, 2, 2, 1, 1, 1, 1) & (4, 3, 3) & 75 &  236 %[250,15]
\\
& (2, 2, 2, 2, 1, 1, 1, 1) & (3, 3, 3, 3) & 60 &  193%[208,16]
\\
\hline
$[5|2 \ 4]$ & (1, 1, 1, 1, 1, 1, 1) & (3, 2, 2) &48 & 159%[175,17]
\\
\hline
$[5|3 \ 3]$ & (1, 1, 1, 1, 1, 1) & (3, 3) & 72 &  213%[216,4]
\\ 
& (1, 1, 1, 1, 1, 1, 1) & (3, 2, 2) & 54 &  166 %[182,17]
\\
& (1, 1, 1, 1, 1, 1, 1, 1) & (2, 2, 2, 2) & 36 &  113 %[128,16]
\\
\hline
$[6|2 \ 2 \ 3]$ & (1, 1, 1, 1, 1, 1, 1) & (3, 2, 2) &72& 213%[231,19]
\\ 
& (1, 1, 1, 1, 1, 1, 1, 1) & (2, 2, 2, 2) & 48 & 145%[160,16]
\\
\hline
\end{tabular}
{\it \caption{The particle content for the $SO(10)$-GUT theories arising from
  our classification of stable, positive, $SU(4)$ monad bundles $V$ on the
  Calabi-Yau threefold $X$. The number $n_{\overline{16}}$ of anti-generations
  vanishes. The number $n_{10}$ vanishes for generic choices of the
  map $g$ in the monad sequence~\eqref{monad}, but can be made
  non-vanishing with particular choices of $g$.}
\label{t:SO10}
}
\end{center}
\end{table}

%+++++++++++++++++++++++++++++++++
\subsection{$SU(5)$-GUT Theories}
Finally, we should consider ${\rm SU}(5)$ GUT theories which originate
from rank $5$ bundles with structure group ${\rm SU}(5)$. To demonstrate
their stability from Hoppe's criterion we have to show that
$H^0(X,\Lambda^pV^*)$ for $p=1,2,3,4$ vanish. For $p=1,2,4$ this
has already been accomplished in Section~\ref{sec:stability}, so it
remains to deal with the case $p=3$.

Unfortunately, for $p=3$ the long exterior power sequences~\eqref{wedge-hom}
together with Kodaira vanishing are not quite sufficient to prove that
$H^0(X,\Lambda^3V^*)=0$. Indeed, writing down~\eqref{wedge-seq} for
$p=3$ we find
\bea
\nn&& 0 \to S^3 C^* \to S^2 C^* \otimes B^* \to K_1 \to 0 \ , \\
&& 0 \to K_1 \to C^* \otimes \wedge^2 B^* \to K_2 \to 0 \ , \label{wedge3split}\\
&& 
0 \to K_{2} \to \wedge^3 B^* \to \wedge^3 V^* \to 0 \ .
\nn
\eea
Now, using the 3 intertwined long exact sequences in cohomology
induced by the above 3 sequences, together with Kodaira vanishing
for the negative bundles formed from the symmetric and
anti-symmetric powers of $B^*$ and $C^*$, we can only conclude that
\beq\label{h0Xwedge3}
H^0(X,\wedge^3 V^*) \simeq H^2(X,K_1) \ .
\eeq
We will now show that the stability proof can be completed by applying Koszul
resolutions to our rank $5$ bundles. This technique makes explicit use of the
embedding in the ambient space ${\cal A} =\mathbb{P}^m$ and its complexity grows with
the number of co-dimensions of the Calabi-Yau manifold $X$ in ${\cal A}$. We, therefore,
start with the quintic, $X=[4|5]$, the only co-dimension one example among the
five Calabi-Yau manifolds under consideration, before we proceed to the more
complicated examples.
\subsubsection{Stability for Rank 5 Bundles on the Quintic}
For the quintic, the normal bundle is simply given by $\cN=\mathcal{O}(5)$
and the Koszul sequence~\eref{koszul}, applied to $\cW=\Lambda^3\cV^*$, explicitly reads
\begin{equation}\label{koszul1}
0\rightarrow \cN^{* }\otimes \wedge ^{3}\cV^{* }\rightarrow \wedge
^{3}\cV^{* }\rightarrow  \wedge ^{3}V^{* }\rightarrow
0 \ .
\end{equation}
From this, we have the long exact sequence in cohomology,
\beq
0\rightarrow H^{0}(\mathcal{A},\cN^{* }\otimes \wedge ^{3}\cV^{*
})\rightarrow H^{0}(\mathcal{A},\wedge ^{3}\cV^{* })\rightarrow
H^{0}(X,\wedge ^{3}V^{* })
\rightarrow H^{1}(\mathcal{A},\cN^{* }\otimes
\wedge ^{3}\cV^{* })\rightarrow ...  
\label{ext}
\eeq
Thus, if we knew $H^{0}(\mathcal{A},\wedge ^{3}\cV^{* })$ and $H^{1}(%
\mathcal{A},N^{* }\otimes \wedge ^{3}\cV^{* })$, we could hope to
determine $H^{0}(X,\wedge ^{3}V^{* })$ itself. In fact, we can show that $%
H^{0}(\mathcal{A},\wedge ^{3}\cV^{* })=H^{1}(\mathcal{A},N^{* }\otimes
\wedge ^{3}\cV^{* })=0$ by writing down the ambient space version of the
exterior power sequences~\eqref{wedge3split} tensored by $\cN^*$.
\begin{eqnarray}
0 &\rightarrow &\cN^{* }\otimes S^{3}\cC^{* }\overset{h}{\rightarrow }
\cN^{* }\otimes S^{2}\cC^{* }\otimes \cB^{* }\rightarrow
\cK_{1}\rightarrow 0  \ , \nn \\
0 &\rightarrow &\cN^{* }\otimes \cK_{1}\rightarrow \cN^{* }\otimes
\cC^{* }\otimes \wedge ^{2}\cB^{* }\rightarrow\cK_{2}\rightarrow 0 
 \ , \\
0 &\rightarrow &\cK_{2}\rightarrow \cN^{* }\otimes \wedge ^{3}\cB^{*
}\rightarrow \cN^{* }\otimes \wedge ^{3}\cV^{* }\rightarrow 0 \
.
\nn
\end{eqnarray}
Since $\cB^{* }$, $\cC^{* }$ and $\cN^{* }$ are all negative bundles, it
follows that $H^{0}(\mathcal{A},\wedge ^{3}\cV^{* })=0$ and 
$h^{1}(\mathcal{A},\cN^{* }\otimes \wedge ^{3}\cV^{* })= h^{3}(\mathcal{A}
,\cK_{1})=\ker (h^{\prime })$, where 
$h^{\prime }:$ $H^{4}({\cal A},\cN^{* }\otimes S^{3}\cC^{* })\rightarrow
H^{4}({\cal A},\cN^{* }\otimes S^{2}\cC^{* }\otimes \cB^{* })$ is the map induced
from $h$ above. Now, we note that since the ranks of the maps in the
defining monads were chosen, by construction, to be maximal rank, 
it follows that the
induced map $h$ in the exterior power sequence is also maximal rank. To
proceed further, we finally observe that for any generic, maximal rank
map $h:{\cal U}\to {\cal W}$ between two ambient space bundles ${\cal U}$ and ${\cal W}$
the induced map $\tilde{h}:H^0({\cal A},{\cal U})\to H^0({\cal A},{\cal W})$
is also maximal rank (see Appendix \ref{ap:h0}).
Since the sequences above are all defined over the ambient space and $h$
is maximal rank, it follows from the above argument that $h^{\prime }$ is
maximal rank and $\ker (h^{\prime })=0$. Therefore, 
\begin{equation}
h^{1}(\mathcal{A},\cN^{* }\otimes \wedge ^{3}\cV^{* })=0.
\end{equation}
Thus, returning to \eref{ext}, we find that $H^{0}(X,\wedge ^{3}V^{* })=0$
and by Hoppe's criterion, \textit{all generic, positive }$SU(5)$\textit{\ bundles are
stable on the quintic}.

\subsubsection{The Co-dimension 2 and 3 Manifolds}

The stability proof for our remaining rank $5$ bundles is similar in
approach, but slightly more lengthy than that given in the previous
subsection. In the
interests of space, we will only give an overview of it here. 
We recall from Subsection \ref{s:class} that the remaining
Calabi-Yau manifolds with rank $5$ bundles are defined by two and three
constraints in $\mathbb{P}^{5}$ and $\mathbb{P}^{6}$ respectively. We first look
at the co-dimension two case.

For co-dimension two, the normal bundle takes the form $N=\mathcal{O}
(q_1)\oplus \mathcal{O}(q_2)$ with $q_1,q_2>0$. This time the Koszul sequence
\eref{koszul} is no longer short-exact, but reads
\begin{equation}
0\rightarrow \wedge ^{2}\cN^{* }\otimes \wedge ^{3}\cV^{* }\rightarrow
\cN^{* }\otimes \wedge ^{3}\cV^{* }\rightarrow \wedge ^{3}\cV^{* }\overset%
{\rho }{\rightarrow }\wedge ^{3}V^{* }\rightarrow 0 \ .
\end{equation}
It can be split into two short exact sequences,
\begin{eqnarray}
0 &\rightarrow &\wedge ^{2}\cN^{* }\otimes \wedge ^{3}\cV^{* }\rightarrow
\cN^{* }\otimes \wedge ^{3}\cV^{* }\rightarrow \cK
\rightarrow 0 \ , \notag \\
0 &\rightarrow &\cK\rightarrow \wedge ^{3}\cV^{* }\overset{
\rho }{\rightarrow }\wedge ^{3}V^{* }\rightarrow 0 \ .
\end{eqnarray}
From the long cohomology sequences of these two resolutions, we find that $
H^{0}(X,\wedge ^{3}V^{* })\simeq$
$H^{2}(\mathcal{A},\wedge ^{2}\cN^{*}\otimes \wedge ^{3}\cV^{* })$
(since $H^{0}(\mathcal{A},\wedge ^{3}\cV^{*
})=H^{0}(\mathcal{A},\cN^{* }\otimes \wedge ^{3}\cV^{* })=0$ by the same
arguments as before). 
Next, the exterior power sequence~\eqref{wedge-seq1},
multiplied by $\Lambda^2\cN^*$ and written over $\mathbb{P}^{5}$ yields,
\begin{eqnarray}
0 &\rightarrow &\wedge ^{2}\cN^{* }\otimes S^{3}\cC^{* }\overset{h}{
\rightarrow }\wedge ^{2}\cN^{* }\otimes S^{2}\cC^{* }\otimes \cB^{*
}\rightarrow \cK_{1}\rightarrow 0  \ , \notag \\
0 &\rightarrow &\cK_{1}\rightarrow \wedge
^{2}\cN^{* }\otimes \cC^{* }\otimes \wedge ^{2}\cB^{* }\rightarrow
\cK_{2}\rightarrow 0  \ , \notag \\
0 &\rightarrow &\cK_{2}\rightarrow \wedge ^{2}\cN^{* }\otimes \wedge
^{3}\cB^{* }\rightarrow \wedge ^{2}\cN^{* }\otimes \wedge ^{3}\cV^{*
}\rightarrow 0 \ .
\end{eqnarray}
Once again, we find that $H^{2}(\mathcal{A},\wedge ^{2}\cN^{* }\otimes
\wedge ^{3}\cV^{* })\simeq H^{4}(\mathcal{A},\cK_{1})$ and $h^{4}(
\mathcal{A},\cK_{1})=\ker (h^{\prime })$ where $h^{\prime }:$ $
H^{5}({\cal A},\wedge ^{2}\cN^{* }\otimes S^{3}\cC^{* })\rightarrow
H^{5}({\cal A},\wedge^{2}\cN^{* }\otimes S^{2}\cC^{* }\otimes \cB^{* })$.
As before, it follows from our definition of the monad that $h^{\prime
}$ is maximal rank
and $\ker (h^{\prime })=0$. Therefore, all positive rank $5$ bundles
on the manifolds $[5|2\;4]$ and $[5|3\;3]$ are stable.

With this analysis complete, we are left with only one rank $5$ bundle
on the co-dimension 3 manifold, $[6|2\;2\;3]$, to consider. In this
case, we could directly apply the Koszul resolution techniques as
above, with a normal bundle, $\mathcal{N}=\mathcal{O}(2) \oplus
\mathcal{O}(2) \oplus \mathcal{O}(3)$, and higher antisymmetric powers
in the Koszul resolution \eref{koszul}. Note, however, that in this
case we are not assured that the dual sequence \eref{monaddual} is
well defined on the ambient space, since the numeric criteria in
Theorem \ref{dualtheorem} are not satisfied on
$\mathbb{P}^6$. However, we can still compute the cohomology of the
relevant sheaves on $\mathbb{P}^6$.  The calculation is lengthy, but
straightforward. 

It is worth noting that there is an alternative approach to this
case. Instead of viewing the Koszul resolution as describing the
restriction of objects on $\mathbb{P}^m$ to the Calabi-Yau, we may
view $X=[6|2\;2\;3]$ as a sub-variety in the 4-fold 
$Y=[6|2\;2]$. Then we may apply the Koszul techniques exactly as before,
viewing the normal bundle to the Calabi-Yau as a line bundle,
$\mathcal{O}_Y(3)$ in $[6|2\;2]$. The analysis then reduces to that
described for the
co-dimension 1 case \eref{koszul1} (that is, that of the rank 5 bundles on the
quintic). A straightforward calculation shows that $H^{0}(X,\wedge
^{3}V^{* })=0$ and the final rank 5 bundle is stable.

\subsubsection{Particle Content}
\begin{table}
\begin{center}
\begin{tabular}{|c|c|c|c|c|c|} \hline
$X$ & $\{b_i\}$ & $\{c_i\}$ & $n_{10}$ & 
 $n_1$\\ \hline
%again [a,b] in comments means [h^1(B^* \otimes V), h^0(C^*\otimes C)]
$[4|5]$ 
 & (1, 1, 1, 1, 1, 1, 1, 1) & (3, 3, 2) & 45 &  202%[216,15]
\\ 
& (1, 1, 1, 1, 1, 1, 1, 1) & (4, 2, 2) &60 & 262%[296,35]
\\ 
& (2, 2, 2, 2, 2, 1, 1, 1, 1, 1) & (3, 3, 3, 3, 3) &75 &301%[325,25]
\\ \hline
$[5|2 \ 4]$
& (1, 1, 1, 1, 1, 1, 1, 1) & (3, 3, 2) & 72 & 288 %[304,17]
\\
\hline
$[5|3 \ 3]$
&(1, 1, 1, 1, 1, 1, 1, 1, 1) & (3, 2, 2, 2) &63 &243%[270,28]
\\ 
&(1, 1, 1, 1, 1, 1, 1, 1, 1, 1) & (2, 2, 2, 2, 2) &45 &176%[200,25]
\\
\hline
$[6|2 \ 2 \ 3]$ & 
(1, 1, 1, 1, 1, 1, 1, 1, 1, 1) & (2, 2, 2, 2, 2) &60 &226%[250,25]
\\
\hline
\end{tabular}
{\it \caption{The particle content for the $SU(5)$-GUT theories arising from
  our classification of stable, positive, $SU(5)$ monad bundles $V$ on the
  Calabi-Yau threefold $X$.
  The number of anti-generations, $n_{\overline 10}$, vanishes. Further,
  $n_5=n_{10}$. Moreover, $n_{\overline{5}}=0$ for generic choices of the map
  $g$ in Eq.~\eqref{monad}, and can be made non-vanishing in special
  regions of moduli space.}
\label{t:SU5}
}
\end{center}
\end{table}
We have shown, using the Koszul sequence, that all positive rank 5
bundles in our classification are stable. Let us now analyze their
particle spectrum. From Table 2, we need to compute $n_{10}=h^{1}(X,V)$,
$n_{\overline{10}}=h^1(X,V^*)= h^{2}(X,V)$, 
$n_{5}=h^{1}(X,\wedge ^{2}V)$, $n_{\overline{5}}=
h^{1}(X,\wedge ^{2}V^*) = h^2(X, \wedge^2V)$, and 
$n_{1}=h^{1}(X,V\otimes V^*)$. As for rank 4 and 5 bundles, we have
$h^0(X,V)=h^3(X,V)=0$ from stability and $h^2(X,V)=2$ from the
sequence~\eqref{dual-seq}. Consequently, we find
\beq
n_{10} = h^1(X,V) =  -\ind(V)\; , \qquad n_{\overline{10}} = 0 \ .
\eeq
As before, we have no anti-generations and the (negative) of the index,
listed in right-most column of Tables 4--7, gives the number $n_{10}$
for all rank $5$ bundles. We include these in \tref{t:SU5} for reference.

Next, we need to compute the $H^1(X,\wedge ^{2}V)$ and $H^2(X,\wedge ^{2}V)$.
From the above arguments we know that $V$ is stable; hence $\wedge
^{2}V$ is also stable and thus $H^0(X,\wedge ^{2}V)$ and $H^3(X,\wedge ^{2}V)$
both vanish (recall that we have already shown explicitly that
$H^0(X,\wedge ^{2}V^*) = H^3(X,\wedge^{2}V)$ vanishes). Therefore, applying
the index theorem~\eqref{index} to $\Lambda^2V$ we have
\beq\label{indwedge2}
-h^1(X,\wedge ^{2}V)+h^2(X,X,\wedge ^{2}V) = \ind(\wedge ^{2}V) =
\frac12 \int_X c_3(\wedge ^{2}V) \ .
\eeq
For $SU(n)$ bundles one has (see Eq.~(339) of Ref.~\cite{Donagi:2004ia}),
\beq\label{c3wedge2}
c_3(\wedge ^{2}V) = (n-4) c_3(V) \ .
\eeq
Hence, combining \eref{indwedge2} and \eref{c3wedge2}, we find the relation
\beq\label{n5-n5bar}
-n_5 + n_{\overline{5}} = \ind(V) = -n_{10} \ .
\eeq
We still need to compute one of the numbers $n_5$ and
$n_{\overline{5}}$. Macaulay~\cite{m2} can very easily
calculate $n_{\overline{5}}= h^{1}(X,\wedge ^{2}V^*) = h^2(X,
\wedge^2V)$. It turns out that
\begin{equation}
 n_{\overline 5}=0
\end{equation}
for all rank $5$ bundles and generic choices~\footnote{Presumably, $n_{\overline 5}$ can
be different from zero for non-generic choices of the map $g$, similar to the case
of $n_{\overline 10}$ for rank $4$ bundles.} of the map $g$. From Eq.~\eqref{n5-n5bar}
this implies
\begin{equation}
 n_5=n_{10}\; ,
\end{equation}
and, hence, the complete spectrum is determined by $n_{10}$ and $n_1$. We have listed
these numbers in~\tref{t:SU5}.

\section{Conclusion}
In this paper, we have presented a classification of positive ${\rm SU}(n)$ monad bundles
on the five Calabi-Yau manifolds defined by complete intersections in a single
projective space. We have required that these bundles can be incorporated
into a consistent heterotic compactification where the heterotic anomaly cancellation
condition can be satisfied by including an appropriately wrapped five-brane.
In addition, we have imposed two ``physical'' conditions, namely that the
rank of bundle be $n=3,4,5$ (in order to obtain a suitable grand unification group)
and that the index of the bundle (that is, the chiral asymmetry) is a non-zero
multiple of three. Given these conditions, we found $37$ bundles on the
five Calabi-Yau manifolds in question, $20$ for rank $3$, $10$ for rank $4$ and
$7$ for rank $5$. 
Using a simple criterion due to Hoppe, we have shown that all
these bundles are stable and, hence, lead to supersymmetric compactifications.
We have also computed the full particle spectrum for all $37$ cases, including
the number of gauge singlets. A generic feature of all our bundles is that
the number of anti-generations vanishes.

These results show that a combination of analytic computations and computer algebra
can be used to analyze a class of models algorithmically. In particular, we have seen
that the notoriously difficult problem of proving stability can be addressed
systematically and that the full particle spectra can be obtained for all cases.
Although the final number of models is still relatively small we expect that
these methods can be extended to much larger classes of Calabi-Yau manifolds, such
as complete intersections in products of projective spaces and in weighted
projective spaces. Such a large-scale analysis which is currently underway~\cite{us}
will lead to a substantial number of examples with broadly the right physical
properties. This class of models can then be used to implement more detailed
particle physics requirements and to systematically search for examples close
to the standard model.

\section*{Acknowledgments}
The authors would like to expression our sincere gratitude to Maria Brambilla,
Philip Candelas, Dan Grayson, Tristan H\"ubsch,
Lionel Mason, Balasz Szendroi and Andreas Wisskirchen for helpful
discussions. Our special thanks go to Adrian Langer whose help with
some of the relevant mathematical problems was invaluable.
%and without
%whom this paper would have not been written in the current form.
L.~A.~thanks the US NSF and the Rhodes Foundation for support. 
Y.-H.~H~is indebted to the FitzJames Fellowship of Merton College, Oxford. 
A.~L.~is supported by the EC 6th Framework Programme
MRTN-CT-2004-503369.  

%%%%%++++++++++++
\appendix
%
%=====
%
\section{Monads, Sheaf Cohomology and Computational Algebraic Geometry} \label{appA}
In this Appendix, we briefly outline some basics of commutative algebra
as relevant for computing sheaf cohomology (see Refs.~\cite{schenck,hart}). In
most computer algebra packages such as Macaulay2 \cite{m2}, 
of which we make extensive use in this paper, these
techniques are essential. Computational algebraic geometry has also been recently used in string phenomenology in \cite{Gray:2006gn} and the reader is referred to tutorials in these papers as well for a quick introduction.
\subsection{The Sheaf-Module Correspondence}
Since we are concerned with compact manifolds, we will focus on projective
varieties in $\IP^m$. A projective algebraic variety is the zero locus of a set of homogeneous
polynomials in $\IP^m$ with coordinates $[x_0:x_1:\ldots:x_m]$.
In the language of commutative algebra, 
projective varieties correspond to homogeneous ideals, $I$, in the polynomial ring
$R_{{\mathbb P}^n} = \IC[x_0, \ldots, x_m]$. An ideal $I\subset R_{{\mathbb P}^n}$,
associated to a variety,
is generated by the defining polynomials of the variety and consists of all
polynomials which vanish on this variety. The quotient ring $A=R_{{\mathbb P}^n}/I$ is called
the {\em coordinate ring} of the variety.

In general, a ring $R$ is called {\it graded} if
\[
R = \bigoplus\limits_{i \in \IZ} R_i, \qquad \mbox{such that }
r_i \in R_i, r_j \in R_j \Rightarrow r_ir_j \in R_{i+j} \; .
\]
For the polynomial ring $R_{{\mathbb P}^n}$ the $R_i$ consists of the homogeneous
polynomials of degree $i$.
In analogy to vector spaces over a field, one can introduce $R$-modules $M$ 
over the ring $R$. In practice, one can think of $M$ as consisting of vectors
with polynomial entries with $R$ acting by polynomial multiplication. 
A module is called {\em graded} if
\[
M =  \bigoplus\limits_{i \in \IZ} M_i, \qquad \mbox{such that }
r_i \in R_i, m_j \in M_j \Rightarrow r_i m_j \in M_{i+j} \ .
\] 
The graded ring $R$ is itself a graded $R$-module, $M(R)$.
Similarly, an ideal $I$ in a graded ring $R$ is a graded $R$-module
and a submodule of $M(R)$. Another important example of a 
graded $R$ module is $R(k)$ which denotes the ring $R$ with
degrees shifted by $-k$. For example, $x^2y \in R_{{\mathbb P}^n}$ is of degree 3, but 
seen as an element of the module $R_{{\mathbb P}^n}(-2)$, its degree is $3+2=5$.

Sheafs over a (projective) variety can also be described as a
module by virtue of the {\it sheaf-module correspondence}.
Given the graded ring $R$ and a finitely generated graded $R$-module
$M$, one defines an associated sheaf $\widetilde{M}$ as follows.
On an open set $U_g$, given by the complement of the zero locus of
$g \in R$, the sections over $U_g$ are
$\widetilde{M}(U_g) = \{ m/g^n |m\in M\, , \mbox{degree}(m) = \mbox{degree}(g^n)\}$.
On $\IP^m$, this looks concretely as follows. A sufficiently fine
open cover of $\IP^m$ is provided by $U_{x_i}$, the open sets where
$x_i\neq 0$. Let us first consider the module $M(R_{{\mathbb P}^n})$, that is, the ring
$R_{{\mathbb P}^n}$ seen as a module. Then $\widetilde{M(R_{{\mathbb P}^n})}(U_{x_i})=
\{f/x_i^m, f \mbox{ homogeneous of degree }n \}$ and, hence,
$\widetilde{M(R_{{\mathbb P}^n})}=\cO_{\IP^m}$, where $\cO_{\IP^m}$ is the trivial
sheaf on $\IP^n$. Similarly, for the modules $R_{{\mathbb P}^n}(k)$ one has
\[
\cO_{\IP^m}(k) \simeq \widetilde{R_{{\mathbb P}^n}(k)} \ .
\]
For projective varieties $X\subset \IP^m$ and associated ideal $I$,
the story is similar. Now, one needs to consider the graded modules over
the coordinate ring $A=R/I$. In particular, for line bundles $\cO_X(k)$
on $X$ one has
\[
 \cO_X(k)=\widetilde{A(k)}\; .
\]
\subsection{Constructing Monads using Computer Algebra}
Recall from \eref{monad2}, that we wish to construct bundles $V$
defined by
\beq\label{monad3}
0 \to V \stackrel{f}{\longrightarrow}
\bigoplus\limits_{i=1}^{r_B} \cO_X(b_i) 
\stackrel{g}{\longrightarrow}
\bigoplus\limits_{i=1}^{r_C} \cO_X(c_i) 
\longrightarrow 0 \ ,
\eeq
over the manifold $X$. In this subsection, we outline how one may
proceed with this construction using commutative algebra packages such
as \cite{m2} and applying the Sheaf-Module correspondence
discussed above. Let $A$ be the coordinate ring of $X$. For example,
for the quintic, $[4|5]$ we can write
\beq
A = \IC[x_0, \ldots, x_4] / \left( \sum_{i=0}^5 x_i^5 + \psi
x_0x_1x_2x_3x_4 \right) \ . \label{Aquintic}
\eeq
where the round brackets denote the ideal generated by the enclosed
polynomial. In practice, we will randomize $\psi$, the complex structure and in
fact work over the ground field $\IZ/p \IZ$ for some large prime $p$
instead of $\IC$ in order to speed up computation. The free modules
corresponding to the bundles $B$, $C$ are given by
$\oplus_{i=1}^{r_B}A(b_i)$, $\oplus_{i=1}^{r_B}A(c_i)$ with grading
$\{b_1, b_2, \ldots, b_{r_B} \}$, $\{c_1, c_2, \ldots, c_{r_C}\}$ and
ranks $r_B$, $r_C$~\footnote{
In most computer packages, the convention is to actually take the
grading to be negative, viz., $\{-b_1, -b_2, \ldots, -b_{r_B} \}$.}.
At the level of modules, the map $g$ can then be specified by an $r_C \times r_B$
matrix whose entries, $g_{ij}$ are homogeneous polynomials of degree $c_i - b_j$,
that is $g_{ij}\in\cO_X(c_i-b_j)$. Indeed, the degrees of the entries of $g$
are so as preserve the gradings of $B$ and $C$ and our choice $c_i \ge b_j$
ensures that such polynomials indeed exist.
Moreover, we choose these polynomials to be random; this corresponds to the
genericity assumption for $g$ used repeatedly in the main text.
%
%\subsection{The Bernstein-Gelfand-Gelfand Correspondence}
\subsection{Algorithms for Sheaf Cohomology}
We shall not delve into the technicalities of this vast subject and
will only mention that for commutative algebra packages such as
\cite{m2}, there are built-in routines for computing
cohomology groups of sheafs (modules). The standard algorithm is based on
the so-called Bernstein-Gel'fand-Gel'fand correspondence and on Tate resolutions
of exterior algebras. The interested reader is referred to the books
\cite{schenck} and \cite{m2book} for details.

\subsection{A Tutorial}\label{ap:eg}
Let us explicitly present a Macaulay2 code \cite{m2}
for one of the examples from our classification. This will serve to illustrate the
power and relative ease with which computer algebra assists in the
proof of stability and the calculation of the particle spectrum.

Let us take the first rank 4 example for $X=[4|5]$ in Table 10,
which was further discussed around Eq.~\eref{5eg-1}. It is defined by
\beq\label{5eg}
B=\cO_X^{\oplus 2}(2) \oplus  \cO_X^{\oplus 4}(1)\; ,\qquad C= \cO_X^{\oplus 2}(4) \ .
\eeq
We work over the polynomial Ring $R_{\IP^4}$ with variables $x_0, \ldots, x_4$
and the ground field $\IZ/27449$. The (projective) coordinate ring
$A$ of a smooth quintic $X$ is then defined following Eq.~\eqref{Aquintic}.
In Macaulay this reads
{\sf \[\bt{rcl}
R &=& ZZ/27449[x\_\{0\}..x\_\{4\}]; \\
A &=& Proj( R/ideal(x\_\{0\}\verb|^|5 + x\_\{1\}\verb|^|5 +
x\_\{2\}\verb|^|5 + x\_\{3\}\verb|^|5 + x\_\{4\}\verb|^|5 + \\
&&2*x\_\{0\}*x\_\{1\}*x\_\{2\}*x\_\{3\}*x\_\{4\})); \\
\et\]}
Next, we define {\sf o}, the trivial sheaf (line-bundle) over $A$, and
the $A$-modules associated to the bundles {\sf B} and {\sf C}.
{\sf \[\bt{l}
o = OO\_(A);\\
B = module (o\verb|^|2 (2) ++ o\verb|^|4 (1)); \\
C = module (o\verb|^|2 (4));\\
\et\]}
Subsequently, a random, generic map, {\sf gmap}, can be constructed
between {\sf B} and {\sf C} (note that in Macaulay, maps are defined
backwards):
\[
\mbox{{\sf gmap = map(C, B, random(C, B));}}
\]
Finally, we can define $V^*$ as the co-kernel of the transpose of {\sf
  fmap}:
\[
\mbox{{\sf Vdual = sheaf coker transpose fmap;}}
\]
We can check that $V^*$ has the expected rank 4 using the command
\[
\mbox{{\sf print rank Vdual;}}
\]
The cohomologies of ${\sf Vdual}$ are easily obtained, for example,
{\sf \[
\mbox{print rank HH\^{~}\!\!2 Vdual;}
\]}
produces 90, precisely as expected. Likewise, one can verify that 
{\sf HH\verb|^|0 Vdual} gives $0$, as is required by stability.
To compute $n_{10}=h^1(X, \wedge^2 V^*)$, one only needs the following
command
\[
\mbox{{\sf print rank HH\^{~}\!\!1 exteriorPower(2, Vdual);}}
\]
which gives 0, as indicated in \tref{t:SO10}. 
For the non-generic map \eref{5eg-2}, one can define
{\sf \[\bt{rcl}
gmap &=& map (C, B, matrix\{\{4*x\_\{3\}\^{~}\!2, 9*x\_\{0\}\^{~}\!2 +
x\_\{2\}\^{~}\!2,  
8*x\_\{2\}\^{~}\!3,\\
&&2*x\_\{3\}\^{~}\!3,4*x\_\{1\}\^{~}\!3,9*x\_\{1\}\^{~}\!3\},
\{x\_\{0\}\^{~}\!2 + 10*x\_\{2\}\^{~}\!2, x\_\{1\}\^{~}\!2,\\
&&9*x\_\{2\}\^{~}\!3, 7*x\_\{3\}\^{~}\!3,9*x\_\{1\}\^{~}\!3 + x\_\{2\}\^{~}\!3,
x\_\{1\}\^{~}\!3 +7*x\_\{4\}\^{~}\!3 \} \} );
\et\]}
One can then check that the cohomologies of $V^*$
remain unchanged with respect to the generic case, that is,
$h^0(X,V^*)=h^1(X,V^*)=h^3(X,V^*)=0$
and $h^2(X,V^*)=90$ while {\sf rank HH\^{~}\!\!1 exteriorPower(2,
Vdual)} now results in $n_{10}=h^1(X,\Lambda^2V^*)=13$.

The singlets are also easy to compute. The group $H^1(X, V \otimes V^*)$ can be thought of as the global Ext-group $Ext^1(V,V) \simeq Ext^1(V^*, V^*)$; this is, again, implemented in \cite{m2}. The command 
``{\sf print rank Ext\^{~}\!\!1(Vdual, Vdual);}'' will give us 277.

\section{Some useful technical results}\label{proofs}

\subsection{Genericity of Maps}\label{ap:gen}
We first state a  helpful fact regarding the genericity of maps in the ambient space.
Consider a morphism $h:\cB \to \cC$ between two sums of line bundles
$\cB =\oplus_{i=1}^{r_B}\cO (b_i)$ and $\cC =\oplus_{i=1}^{r_C}\cO (c_i)$ on ${\cal A} ={\mathbb P}^m$.
The map $h$ can explicitly be specified by a $r_C\times r_B$ matrix $h_{ij}\in \cO (c_i-b_j)$
and it induces a map $\tilde{h}:H^0({\cal A},\cB )\rightarrow H^0({\cal A} ,\cC )$.
The induced map $\tilde{h}$ is also described by $h_{ij}$ acting on the sections
of $\cB$ and, hence, if the matrix $(h_{ij})$ has maximal rank (almost everywhere) then
$\tilde{h}$ has maximal rank.

\subsection{Proof of $H^0(X,V)=0$}\label{ap:h0}
In this section we will provide a proof that $H^0(X,V)=0$ for all the
bundles defined by positive monads on cyclic complete intersection
Calabi-Yau manifolds. The proof is similar in spirit to the stability
proof of Section 4.5 in that we approach the problem from the point of
view of an embedding space and use Koszul sequences \eref{koszul} to
determine the necessary cohomology.

Because the Koszul resolutions depend on the normal bundle to the
Calabi-Yau, the length of the calculation increases with the
co-dimension of the embedded Calabi-Yau. For conciseness, we will only
provide the proof for the co-dimension 1 case (the quintic) here,
however the higher co-dimension cases follow by an entirely analogous
construction. 
Consider a positive bundle defined by \eref{monad} on the quintic
($X=[4|5]$). Clearly, the normal bundle is simply
$\cN=\mathcal{O}(5)$, and from \eref{koszul} we just obtain the short
exact sequence,
\beq\label{shortkoszul}
0\rightarrow \mathcal{N}^*\otimes \mathcal{V}\rightarrow
\mathcal{V}\rightarrow  V\rightarrow 0 \ .
\eeq
As before, we have the long exact sequence in cohomology,
\beq
0\rightarrow H^{0}(\mathcal{A},\mathcal{N}^{* }\otimes
\mathcal{V})\rightarrow H^{0}(\mathcal{A},\mathcal{V})\rightarrow 
H^{0}(X,V )\rightarrow H^{1}(\mathcal{A},\mathcal{N}^{* }\otimes
\mathcal{V})\rightarrow ...
\eeq 
So, in order to compute $H^{0}(X,V)$, we must first find
$H^{0}(\mathcal{A},\mathcal{V})$ and
$H^{1}(\mathcal{A},\mathcal{N}^*\otimes \mathcal{V})$. To do this, we
will define the following short exact sequences on the ambient space:
\beq\label{ambient1}
0 \rightarrow \mathcal{V} \rightarrow \mathcal{B}
\stackrel{g}\rightarrow \mathcal{C} \rightarrow 0 \ ,
\eeq
and the same sequence tensored with the dual of the normal bundle,
\beq\label{ambient2}
0 \rightarrow \mathcal{N}^*\otimes \mathcal{V} \rightarrow
\mathcal{N}^*\otimes B \stackrel{h}\rightarrow \mathcal{N}^*\otimes
\mathcal{C} \rightarrow 0 \ . 
\eeq
From the Bott Vanishing formula \cite{hart} we have the following
formula for the cohomology of line bundles on the ambient space.
\beq\label{bottformula}
h^{q}(\mathbb{P}^{n}, \mathcal{O}_{\IP^n}(k))=\left\{
\begin{array}
[c]{ll}%
\binom{k+n}{n} & q=0\quad k>-1\\
1 & q=n\quad k=-n-1\\
\binom{-k-1}{-k-n-1} & q=n\quad k<-n-1\\
0 & \mbox{otherwise}
\end{array}
\right. \ .
\eeq
Using this to compute elements of the various long exact cohomology sequences corresponding to \eqref{ambient1} and \eqref{ambient2} we find the following
\bea
&&0\rightarrow H^{0}(\mathcal{A}, \mathcal{V})\rightarrow H^{0}(\mathcal{A},\mathcal{B})
\stackrel{g'}\rightarrow
H^{0}(\mathcal{A},\mathcal{C})\rightarrow H^{1}(\mathcal{A},\mathcal{V})\rightarrow 0 \ , \\
\nn&&0\rightarrow H^{0}(\mathcal{A}, \mathcal{N}^*\otimes \mathcal{V})\rightarrow H^{0}(\mathcal{A},\mathcal{N}^*\otimes \mathcal{B})\stackrel{h'}\rightarrow
H^{0}(\mathcal{A},\mathcal{N}^*\otimes \mathcal{C})\rightarrow H^{1}(\mathcal{A},\mathcal{N}^*\otimes \mathcal{V})\rightarrow 0 \ .
\eea
Now, we note that since the maps in \eref{ambient1} and
\eref{ambient2} were chosen to be maximal rank on $\mathcal{A}$ (that
is, the sequences were constructed to be exact) it follows from the
arguments in Appendix \ref{ap:gen} that the induced cohomology maps
$g'$ and $h'$ on the ambient space are also maximal rank. With these
results in hand, we then find that 
\bea
\nn&&h^0(\mathbb{P}^n,\mathcal{V})=h^0(\mathbb{P}^n, \mathcal{B})-{\rm
  rk}(g') \ , \\
&&h^1(\mathbb{P}^n, \mathcal{N}^*\otimes
\mathcal{V})=h^0(\mathbb{P}^n, \mathcal{N}^*\otimes \mathcal{C})-{\rm
  rk}(h') \ .
\eea

Clearly, we will have $h^0(\mathbb{P}^n,\mathcal{V})=0$ and
$h^1(\mathbb{P}^n, \mathcal{N}^*\otimes \mathcal{V})=0$ if $g'$ and
$h'$ are maximal rank and if
\bea
\nn&&h^0(\mathbb{P}^n,\mathcal{B})\leq h^0(\mathbb{P}^n,\mathcal{C}) \
, \\
\label{ambientcond}&&h^1(\mathbb{P}^n, \mathcal{N}^*\otimes
\mathcal{C})\leq h^1(\mathbb{P}^n, \mathcal{N}^*\otimes \mathcal{B}) \
.
\eea
However, using \eref{bottformula} and the defining sums of line bundles on the ambient space \eref{monad} we find by direct calculation that \eref{ambientcond} is satisfied for all the rank 4 bundles on the quintic. Therefore, we find that $h^0(X,V)=0$ for all rank 4 bundles on the quintic.

The analysis for all the other ranks is similar in construction (with
additional wedge powers of the normal bundle in \eref{shortkoszul}) as
long as Theorem \ref{dualtheorem} is satisfied and we can consistently
define our monads on the ambient space. That is, we must be able to
write the short exact sequences \eref{monad} on the ambient space with
maps whose degeneracy loci vanish. Thus, the technique above can only
be applied directly to rank 4 and 5 bundles on $\mathbb{P}^4$ and rank
5 bundles on $\mathbb{P}^5$. For all the other rank 4 and 5 bundles in
the list, we again apply the techniques of Koszul sequences, but to a
4-fold in the embedding space rather than $\mathbb{P}^m$ itself. For
the rank 3 bundles in our list we cannot apply these methods, nor do
we make use of $h^0(X,V)=0$ in those stability proofs, however we can
verify that the identity holds for all the rank 3 bundles as well by
computer calculation. Thus, we find that $h^0(X,V)=0$ for all the
bundles in our list and verify this by computer algebra using
\cite{m2}.
\subsection{Proof that $n_1=h^1(X,\wedge^2 V^*)=0$ for the $SO(10)$
  Models}\label{SO(10)proof}
For simplicity, we provide here the argument for rank 4 bundles on the
quintic. As in the previous discussion, the proof is easily extended
to the other cases. We begin once again with the Koszul sequence in
the co-dimension 1 case, this time for $\wedge^2 V^*$:
\beq
0\rightarrow \cN^{* }\otimes \wedge ^{2}\cV^* \rightarrow \wedge
^{2}\cV^* \rightarrow  \wedge ^{2}V^* \rightarrow
0 \ .
\eeq
From this, we have the long exact sequence in cohomology,
\beq
...\rightarrow H^{1}(\mathcal{A},\cN^{* }\otimes \wedge ^{2}\cV^* )
\rightarrow H^{1}(\mathcal{A},\wedge ^{2}\cV^* )\rightarrow
H^{1}(X,\wedge ^{2}V^* )
\rightarrow H^{2}(\mathcal{A},\cN^{* }\otimes
\wedge ^{2}\cV^* )\rightarrow ...\  
\eeq
We will show that $h^{1}(X,\wedge ^{2}V^* )=0$ by proving that
$h^{1}(\mathcal{A},\wedge ^{2}\cV^* )$ and $h^{2}(\mathcal{A},\cN^{* }\otimes
\wedge ^{2}\cV^* )$ both vanish.

We begin with $h^{1}(\mathcal{A},\wedge ^{2}\cV^* )$. To proceed, we
have the exterior power sequences  
\bea
&& 0\to S^2\mathcal{C}^*\to\mathcal{C}^*\otimes
 \mathcal{B}^*\to\Lambda^2\mathcal{B}^*\to\Lambda^2
 \mathcal{V}^*\to 0 \ , \\
&&0\to \mathcal{N}^* \otimes S^2\mathcal{C}^*\to \mathcal{N}^*
\otimes \mathcal{C}^*\otimes \mathcal{B}^*\to \mathcal{N}^*
\otimes \Lambda^2 \mathcal{B}^*\to \mathcal{N}^* \otimes \Lambda^2
\mathcal{V}^*\to 0\; , 
\eea
which we can split into the short exact sequences:
\bea
0\to S^2\mathcal{C}^*\to \mathcal{C}^*\otimes
\mathcal{B}^*\to K_1\to 0 \ , \nn \\ 
0\to K_1\to \Lambda^2 \mathcal{B}^*\to\Lambda^2 \mathcal{V}^*\to
0 \ , 
\label{first}
\eea
and similarly,
\bea
0\to \mathcal{N}^* \otimes S^2\mathcal{C}^*\to \mathcal{N}^*
\otimes \mathcal{C}^*\otimes \mathcal{B}^*\to K_2\to 0 \ , \nn\\
0\to K_2\to \mathcal{N}^* \otimes \Lambda^2 \mathcal{B}^*\to {N}^*
\otimes \Lambda^2 \mathcal{V}^*\to 0\; .\ 
\label{second}
\eea
Each of these generates a long exact sequence in cohomology. Using the
familiar results for the cohomologies of positive and negative line
bundles on the ambient space, from \eqref{first} we immediately obtain
$h^1(\mathcal{A}, \Lambda^2 \mathcal{V}^*)=h^2(K_1)=0$. Likewise,
the cohomology sequence of \eqref{second} leads us to
$h^2(\mathcal{A}, \mathcal{N}^* \otimes\Lambda^2
\mathcal{V}^*)=h^3(K_2)=0$ and
\bea
0\to H^3(\mathcal{A},K_2)\to H^4(\mathcal{A},\mathcal{N}^* \otimes S^2
\mathcal{C}^* )\stackrel{f}{\longrightarrow}
H^4(\mathcal{A},\mathcal{N}^* \otimes\mathcal{C}^*\otimes
\mathcal{B}^* )\to H^4(\mathcal{A},K_2) \to 0 \ . \
\eea
Combining these results we find
\bea
h^2(\mathcal{A}, \mathcal{N}^* \otimes
\wedge^2\mathcal{V}^*)=h^4(\mathcal{A}, \mathcal{N}^* \otimes S^2
\mathcal{C}^*)-{\rm rk}(f) \ .
\eea
Now, as before we note that by maximal rank arguments of \ref{ap:gen}
and Serre duality, 
\beq
{\rm rk}(f)={\rm min}(h^4(\mathcal{A}, \mathcal{N}^* \otimes S^2
\mathcal{C}^*), h^4(\mathcal{A},\mathcal{N}^*
\otimes\mathcal{C}^*\otimes \mathcal{B}^* ))
\eeq 
By direct computation using \cite{m2} we find that $h^4(\mathcal{A},
\mathcal{N}^* \otimes S^2 \mathcal{C}^*)<
h^4(\mathcal{A},\mathcal{N}^* \otimes\mathcal{C}^*\otimes
\mathcal{B}^* )$ for all the bundles in our list. Thus,
$h^2(\mathcal{A}, \mathcal{N}^* \otimes \wedge^2\mathcal{V}^*)=0$ and
we may conclude that
\beq
h^1(\mathcal{A}, \wedge^2\mathcal{V}^*)=0 \ .
\eeq
The argument is the same in spirit for the other manifolds in our
list. The only key difference being the length of the starting Koszul
sequence (which will containing higher wedge powers of
$\mathcal{N}^*$). The resulting cohomology analysis follows
straightforwardly.

%
%=++++++++++++++++++++++++++++++++++++++++++++++++++++++++++
%

\end{document}